\documentclass[sn-nature,onecolumn]{sn-jnl}

\usepackage[english]{babel}
\usepackage[utf8]{inputenc}
\usepackage{graphicx}
\usepackage{amsmath,amssymb,amsfonts}
\usepackage{xcolor}
\usepackage{manyfoot}
\usepackage{siunitx}
\newcommand{\pav}[1]{\textcolor{black}{#1}}
\usepackage{geometry}
\usepackage{caption}
\usepackage{setspace}
\begin{document}

\title{Incipient nematicity from electron flat bands in a kagomé metal} 

\author[1,2]{\fnm{Nathan C.} \sur{Drucker}}\email{ndrucker@g.harvard.edu}
\author[1,3]{\fnm{Thanh} \sur{Nguyen}}
\author[1,3]{\fnm{Manasi} \sur{Mandal}}
\author[1,4]{\fnm{Phum} \sur{Siriviboon}}
\author[5]{\fnm{Yujie} \sur{Quan}}
\author[1,3]{\fnm{Artittaya} \sur{Boonkird}}

\author[1,6]{\fnm{Ryotaro} \sur{Okabe}}
\author[7]{\fnm{Fankang} \sur{Li}}
\author[7]{\fnm{Kaleb} \sur{Burrage}}
\author[7]{\fnm{Fumiaki} \sur{Funuma}}
\author[7]{\fnm{Masaaki} \sur{Matsuda}}
\author[7]{\fnm{Douglas L.} \sur{Abernathy}}
\author[7]{\fnm{Travis J.} \sur{Williams}}
\author[7]{\fnm{Songxue} \sur{Chi}}
\author[7]{\fnm{Feng} \sur{Ye}}
\author[8]{\fnm{Christie S.} \sur{Nelson}}
\author[5]{\fnm{Bolin} \sur{Liao}}
\author[9,10]{\fnm{Pavel} \sur{Volkov}}\email{pavel.volkov@uconn.edu}
\author[1,2]{\fnm{Mingda} \sur{Li}}\email{mingda@mit.edu}

\affil[1]{\orgdiv{Quantum Measurement Group}, \orgname{MIT}, \orgaddress{\city{Cambridge}, \state{MA} \postcode{02139}, \country{USA}}}
\affil[2]{\orgdiv{John A. Paulson School of Engineering and Applied Sciences}, \orgname{Harvard University}, \orgaddress{\city{Cambridge}, \state{MA} \postcode{02138}, \country{USA}}}
\affil[3]{\orgdiv{Department of Nuclear Science and Engineering}, \orgname{MIT}, \orgaddress{\city{Cambridge}, \state{MA} \postcode{02139}, \country{USA}}}
\affil[4]{\orgdiv{Department of Physics}, \orgname{MIT}, \orgaddress{\city{Cambridge}, \state{MA} \postcode{02139}, \country{USA}}}

\affil[5]{\orgdiv{Department of Mechanical Engineering}, \orgname{University of California, Santa Barbara}, \orgaddress{\city{Santa Barbara}, \state{CA} \postcode{93106}, \country{USA}}}

\affil[7]{\orgdiv{Department of Chemistry}, \orgname{MIT}, \orgaddress{\city{Cambridge}, \state{MA} \postcode{02139}, \country{USA}}}
\affil[7]{\orgdiv{Neutron Scattering Division}, \orgname{Oak Ridge National Laboratory}, \orgaddress{\city{Oak Ridge}, \state{TN} \postcode{37830}, \country{USA}}}
\affil[8]{\orgdiv{X-ray Scattering Division, NSLS-II}, \orgname{Brookhaven National Laboratory}, \orgaddress{\city{Upton}, \state{NY} \postcode{11973}, \country{USA}}}
\affil[9]{\orgdiv{Deparment of Physics}, \orgname{Harvard University}, \orgaddress{\city{Cambridge}, \state{MA} \postcode{02138}, \country{USA}}}
\affil[10]{\orgdiv{Deparment of Physics}, \orgname{University of Connecticut}, \orgaddress{\city{Storrs}, \state{CT} \postcode{06269}, \country{USA}}}

\abstract{Engineering new quantum phases \pav{often} requires fine tuning of the electronic, orbital, spin, and lattice degrees of freedom. To this end, kagomé lattice with flat bands has garnered great attention by hosting various topological and correlated phases,\pav{ when the flat band is at the Fermi level.} Here we discover the unconventional nematicity in kagomé metal CoSn\pav{, where flat bands are fully occupied below the Fermi level}. Thermodynamic, dilatometry, resonant X-ray scattering, inelastic neutron scattering, Larmor diffraction, and thermoelectric measurements consistently hint \pav{at} rotational symmetry-breaking and nematic order that is pronounced only near T$^* = 225$ K. The observations, principally the nematic's finite temperature stability--incipience-- can be explained by a phenomenological model which reveals that thermally excited flat bands promote symmetry breaking at a characteristic temperature.  Our work shows that thermal fluctuations, which are typically detrimental for correlated electronic phases, can induce new ordered states of matter\pav{, avoiding the requirements for fine tuning of electronic bands}. }

\maketitle

\begin{figure*}[ht!]
 \centering
 \includegraphics[width=\textwidth]{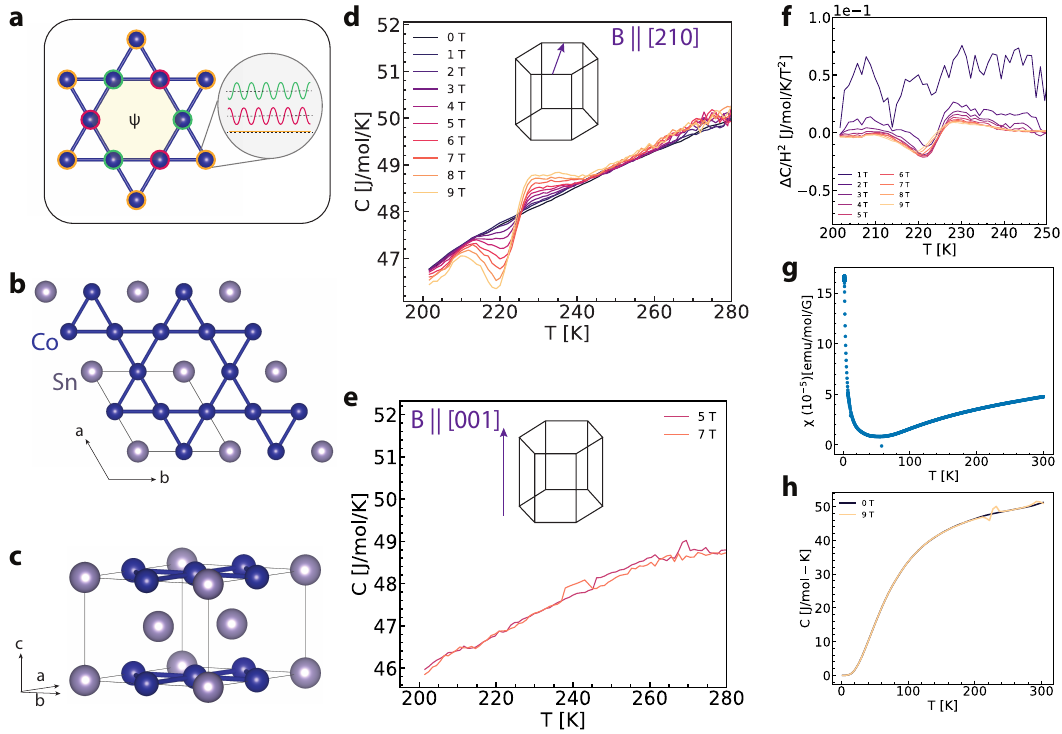}
 \caption{\textbf{Field-induced heat capacity anomaly in geometrically frustrated CoSn.} \textbf{(a)} The kagomé lattice localizes modes when wavefunctions $\psi$ have different phases on alternating corners of the hexagon in the lattice (red and purple, respectively). They destructively interfere on the corner-shared triangles (green). This geometric frustration causes modes to become localized within hexagonal plaquettes. \textbf{(b,c)} CoSn hosts a kagomé net of Co (Blue) atoms in the $a$-$b$ plane separated by honeycomb layers of Sn (grey) atoms. \textbf{(d)} Temperature-dependent heat capacity of CoSn measured at finite magnetic fields applied within the kagomé plane, along the $[210]$ direction. \textbf{(e)} When the field is applied along the $c$-axis, there is no field-induced heat capacity anomaly. \textbf{(f)} For different magnetic field strengths $B$, the heat capacity is shown to scale with the square of the field $B^2$ at fields above B $= 2$ T. \textbf{(g)} Magnetic susceptibility for B $=5000$ Oe applied in the kagomé plane does not show any anomalies. \textbf{(h)} Heat capacity between T $=2$ K and T $=300$ K and B $=0$ T and B $=9$ T. }
 \label{Fig1}
\end{figure*}

Competition between classical and quantum mechanical effects is at the heart of many emergent behaviors in quantum materials which have significant technological impact such as magnetism and superconductivity. A paradigmatic example is the competition between the delocalizing effects of quantum kinetic energy and the Coulomb repulsion which favors classical, point-like particle states of the electron. A fascinating complexity of behaviors is expected to arise in systems where these effects are closely competing, \pav{epitomized by the Hubbard model \cite{hubbard_rev}.} 
This idea has motivated the synthesis of new bulk materials \cite{Kang2020, Kang2020b, Ye2021, Regnault2022, Wakefield2023}, 2D heterostructures \cite{Cao2018, Cao2018a, Li2018}, and artificial superlattices  \cite{Wu2007,Kollar2019a, Kollar2020, Samajdar2021} which aim to reduce kinetic energy, thus enhancing the effects of interactions. An example of this approach are \pav{kagomé materials \cite{neupert2022charge,fernandes2023kagomé,Wang2023_rev,checkelsky2023flat}}, where subtle quantum interference effects lead to a complete quenching of the kinetic energy of electrons, which form localized electronic states but away from atomic centers (Fig. \ref{Fig1}a). \pav{Interaction effects in this case facilitate formation of numerous unconventional phases at low temperatures when such flat bands are partially filled.} Within the Mott-Hubbard paradigm, adding the effects of temperature or diluting the electron concentrations are usually detrimental to the strong correlation effects. As a consequence of this reasoning, quantum materials are often designed with the goal of having electrons with quenched kinetic energy--corresponding to flat bands in reciprocal space--directly at the Fermi-level. 

An archetypal example of this family of materials is CoSn, where the flat band is one of the closest to the Fermi level in known materials, but no phase transitions or magnetic orderings have been reported so far \cite{Kang2020b, Liu2020, Meier2020, Huang2022, Wan2022Temp}. CoSn has a crystal structure composed of 2D sheets of Co atoms arranged in a kagomé lattice separated by layers of Sn atoms arranged in a honeycomb lattice (Fig. \ref{Fig1}b,d). The nearly ideal kagomé lattice of Co atoms causes flat electronic bands, as highlighted through angle-resolved photoemission (ARPES) measurements and density functional theory (DFT) calculations of the electronic band \pav{structure \cite{Kang2020b}.} CoSn does not order magnetically, \pav{unlike} similar compounds FeSn \cite{Kang2020}, Mn$_3$Sn \cite{Kuroda2017}, Fe$_3$Sn$_2$ \cite{Ye2018}, and FeGe \cite{Yin2022FeGe, Teng2022, Teng2023}. From the Mott-Hubbard point of view, the lack of ordering in CoSn is expected, since the flat band is not at the Fermi level. However, magnetic susceptibility measurements indicate that increasing temperature results in a prominent activation of carriers in the flat band, raising the question of the behavior of such electrons in the non-degenerate limit \cite{Meier2020}.

Here, we demonstrate another way in which emergent, interaction-driven behavior can be realized in flat band materials. In particular, we show that non-degenerate, thermally activated \pav{holes} from the flat band in CoSn drive a profound reconstruction of the state of the crystal, making it anomalously prone to develop symmetry-breaking (nematic) deformations. We combine thermodynamic (specific heat, magnetostriction), structural (Larmor diffraction, inelastic neutron scattering), and spectroscopic (resonant-elastic X-ray scattering) probes to unambiguously reveal the presence of rotational symmetry breaking in a narrow interval of temperatures around T* $\sim 225$ K in CoSn. Pronounced field-induced heat capacity and thermopower anomalies in the same temperature range point to intertwined effects of electrons, sensitive to magnetic fields, and lattice, carrying most of the entropy.  We attribute the anomalous behaviors to the effects from thermally activated carriers in the flat kagomé band, which drive the system to the brink of an electronic nematic transition, making the material anomalously susceptible to symmetry-breaking effects of magnetic field and structural distortions. 

Our work demonstrates another facet of emergent behavior arising from quantum-classical competition; rather than restoring symmetry, classical effects of thermal fluctuation of flat bands can reduce the symmetry and lead to anomalous emergent behavior when competing with the quantum interference effects that leads to the electronic flat band. The effects we observe occur in the limit of dilute concentration of thermally activated, non-degenerate \pav{carriers}, where the Mott-Hubbard paradigm does not operate, pointing towards a new pathway to realize new correlated electron systems by engineering flat bands away from the Fermi level.

\begin{figure*}[ht!]
  \centering
  \includegraphics[width=0.75\linewidth]{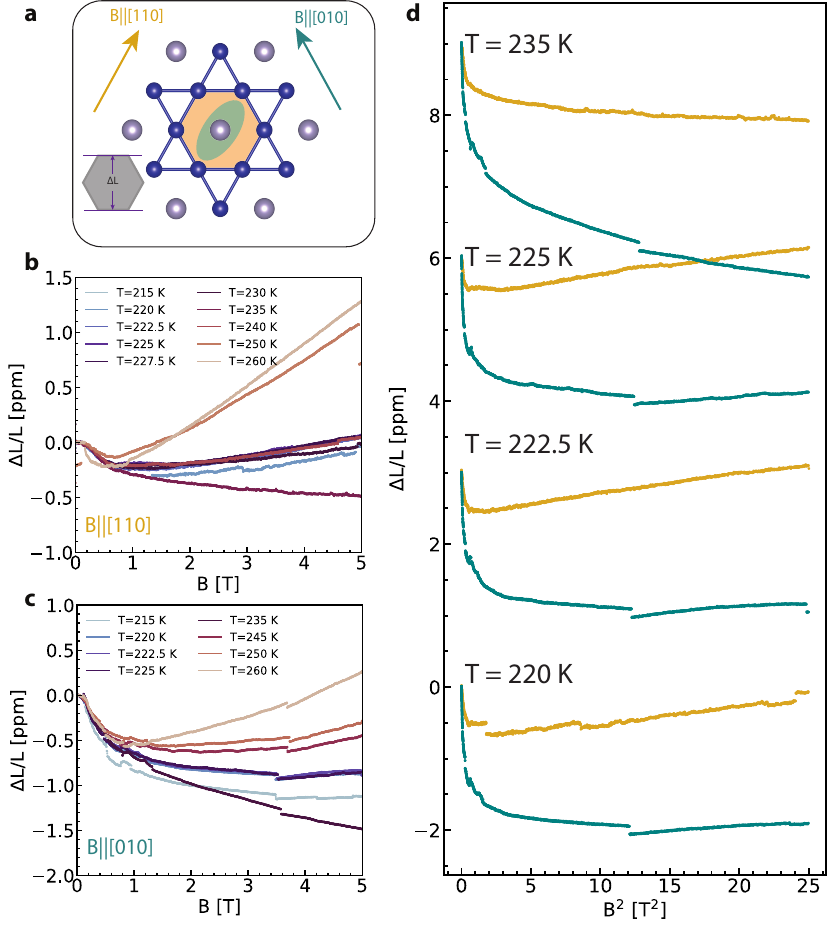}
 \caption{\textbf{Anisotropic magnetostriction within the kagomé plane.} \textbf{(a)} Schematic of magnetostriction experiment with magnetic field applied along different crystallographic axes (purple and green arrows). The turquoise ellipse indicates a possible orientation of the nematicity relative to field directions. Inset in bottom left shows the direction of the measured dilation. \textbf{(b)} Sample dilation along the [210] direction at temperatures near the anomaly at T $=225$ K with magnetic field applied along the [110] direction. \textbf{c} Same as (b) but with magnetic field applied along the [010] direction. \textbf{(d} Dilation as a function of $B^2$ for fields applied along the same directions as \textbf{(b, c)}, respectively. For fields applied along the [110] direction, the magnetostriction scales with the square of the field, between 220 K and 230 K, whereas when the field is applied along the [010] direction, the magnetostriction plateaus at high field. At higher temperatures for the field along the [010] direction, the $B^2$ scaling is restored.}
 \label{Fig2}
\end{figure*}


\section*{Results}
\subsection*{Thermodynamic anomalies}
We begin by presenting thermodynamic evidence for anomalous behavior in CoSn. In Fig. \ref{Fig1}h, we show the heat capacity of CoSn with a magnetic field applied in the kagomé plane. The application of magnetic field results in a distinct anomaly developing at T$^*$ $\sim 225$ K. The data taken near T$^*$ with higher resolution is shown in Fig. \ref{Fig1}d. One observes that the anomaly develops continuously with magnetic field and is almost absent at zero field (see Supplemental Materials (SM) for additional discussion). In particular, the data above B $=2$ T (Fig. \ref{Fig1}f) scales with $B^2$. In contrast, the anomalous behavior is absent for magnetic fields applied perpendicular to the kagomé plane (Fig. \ref{Fig1}e). 

Contrary to the initial expectation that this behavior may come from magnetic ordering, the magnetic susceptibility of CoSn (Fig. \ref{Fig1}g) shows a usual behavior at all temperatures, consistent with prior studies \cite{Meier2020, Huang2022}. This rules out a magnetic transition as the origin of the field-induced heat capacity anomaly. Furthermore, the anomalous orbital magnetization of the electronic flat bands \cite{Yin2019css, Huang2022} in CoSn found for $H\parallel c$ \cite{Huang2022} cannot explain the anomaly appearing for $H\parallel ab$.

However, apart from orienting the magnetic moments, magnetic field also reduces the spatial symmetry of the system ($D_{6h}$ point group). In this respect, the symmetry properties of in-plane and out-of-plane fields are rather different. For $\textbf{B}=(B_x,B_y,B_z)$, $B_z$ transforms as a $A_{2g}$ representation while $B_{x,y}$ transform under $E_{2g}$ representations. Of these, only the in-plane field $B_{x,y}$ component can actually reduce the crystalline symmetry\pav{, e.g. by coupling quadratically to non-symmetric strain}, which appears to agree with the observed anisotropy of the field-induced heat capacity effect in Fig. \ref{Fig1}. 
To verify the connection between the in-plane magnetic field and breaking of the crystalline symmetry, we perform magnetostriction measurements along high-symmetry crystallographic directions. In this experiment, the change in length of the sample is measured between the faces of the hexagonal crystal in the [120] direction, as depicted in Fig. \ref{Fig2}a. To account for possible strain induced by the experimental setup, the magnetic field is applied symmetrically about this direction along the [010] and [110] directions. The dilation for fields applied along the [110] and [010] directions at temperatures near the field-induced heat capacity anomaly as a function of magnetic field are plotted in Fig. \ref{Fig2}b and \ref{Fig2}c, respectively.
At high fields, the elongations exhibits $\propto B^2$ dependence (Fig. \ref{Fig2}c) consistent with the absence of magnetism (note that there are deviations from this scaling behavior at low fields similar to the specific heat data displayed in Fig. \ref{Fig1}). The sign of the effect changes with temperature -- at higher temperatures, the slope is positive, while turning negative at T $\sim$ 235 K. We note that in isotropic weakly correlated metals, magnetostriction is typically positive \cite{chandrasekhar1971magnetostriction}.


Most importantly, we observe that the magnetostriction is different for [110] and [010] field orientations, which are related by a mirror symmetry $\sigma_d$. Such behavior is clearly inconsistent with the full $D_{6h}$ symmetry of the kagomé layers. \pav{In particular, the breaking of $\sigma_d$ is consistent with an order parameter of $E_{2g}$ representation, such as strain $\varepsilon_{xy}$ or electronic nematic order. Presence of such order is also consistent with a $\propto B^2$ contribution to the specific heat for ${\bf B}\parallel ab$ only, since $\{B_x^2-B_y^2, 2B_xB_y\}$ also transforms as $E_{2g}$.}



Observation of these thermodynamic anomalies points to \pav{reduction of symmetry} in CoSn at temperatures close to T$^*$. They also point to the intertwined roles of the lattice and electrons in this phenomenon. The sensitivity to magnetic field is unlikely to come from lattice effects alone. On the other hand, the strength of field-induced heat capacity anomaly surpasses an estimate of total electronic specific heat at T$^*$, obtained by extrapolating the $C\propto T$ term at low temperatures (see SM), indicating that the lattice degrees of freedom are strongly affected. Nevertheless, the electronic and lattice \pav{contributions} ultimately enter thermodynamic quantities on the same footing, precluding the possibility \pav{to disentangle their contributions} . To clarify the origin and mechanism of the observed \pav{symmetry breaking} we thus have to turn to other \pav{symmetry-sensitive} methods, which can distinguish between electronic and lattice degrees of freedom.




\begin{figure*}[ht!]
 \centering
 \includegraphics[height=0.65\textheight]{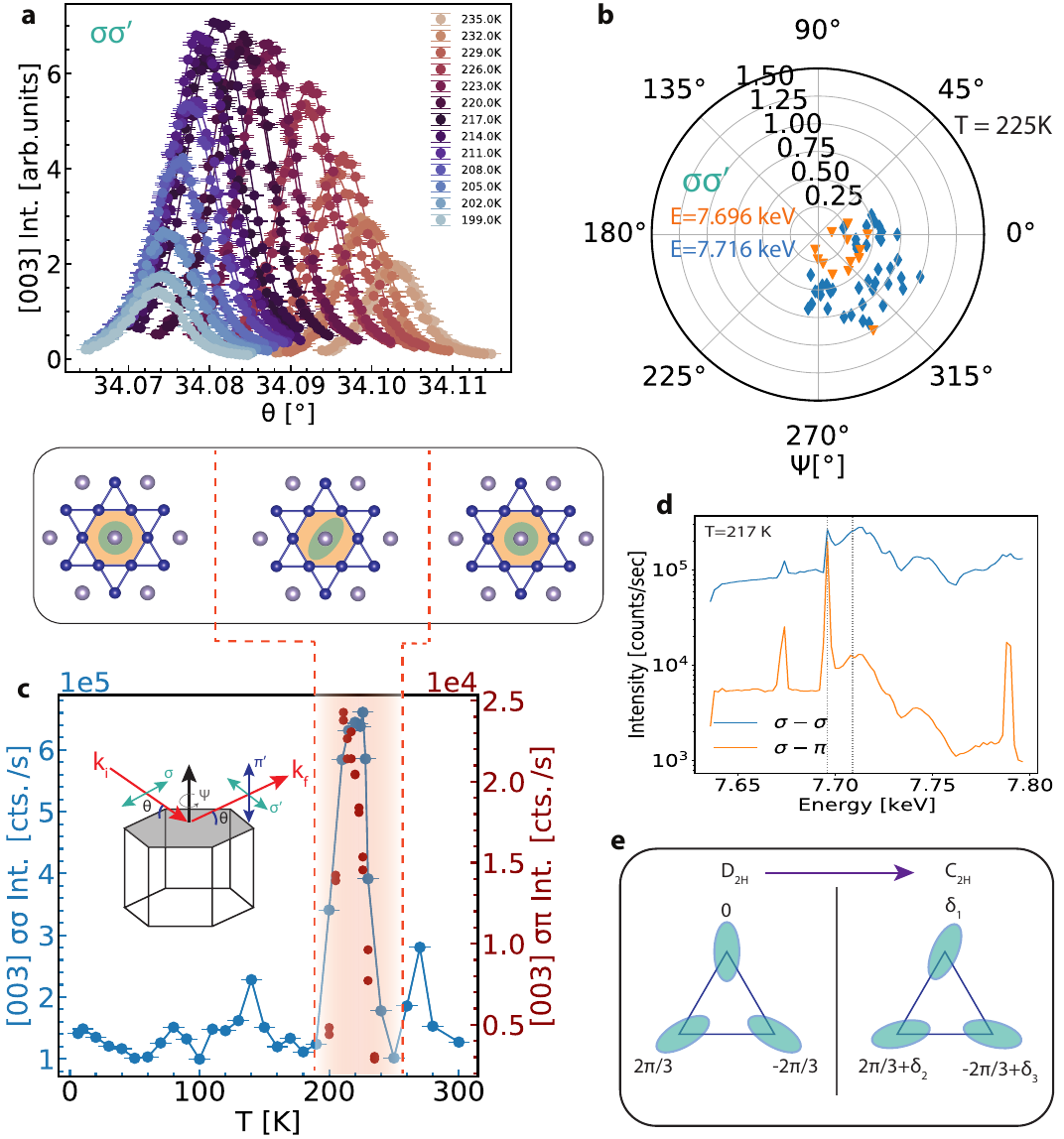}
 \caption{\textbf{Rotational symmetry breaking revealed through REXS at the Co K-edge.} \textbf{(a)} Experimental rocking curves of the [003] Bragg peak show an increase in intensity in the same temperature region as the field-induced heat capacity anomaly and magnetostriction experiments. The incident photon energy is tuned to E $=7.709$ keV to be resonant with the Co K-edge, and the polarization is set in the $\sigma-\sigma'$ channel. \textbf{(b)} Polar plot of azimuthal scans in the $\sigma-\sigma'$ polarization channel taken at T $\sim 225$ K and photon energies E $=7.696$ keV and E $=7.709$ keV in orange and blue, respectively. \textbf{(c)} Integrated intensity of the [003] Bragg peak as a function of temperature in both the $\sigma-\sigma'$ and $\sigma-\pi'$ polarization channels plotted in blue and red, respectively. Orange dashed lines and highlighted region indicate the temperature range where nematicity has been observed in other measurements. Inset shows the scattering geometry, with incident and scattered photon momenta and polarizations labeled, along with scattering angle $\theta$ and azimuthal orientation $\psi$. \textbf{(d)} Energy dependence of the [003] scattering intensity near the Co K-edge for two different polarization channels. Grey dashed lines indicate the energies selected for the experiments. \textbf{(e)} Schematic of anisotropic charge distributions at the Co site. When the symmetry is reduced from $D_{2H}$ to $C_{2H}$, the rotational symmetry of the system is broken and an azimuthal dependence of the scattering intensity in the $\sigma-\sigma'$ channel is allowed by the new symmetry. }
\label{Fig3}
\end{figure*}

\subsection*{Probing electronic nematicity with REXS}
To investigate the role of electronic degrees of freedom in this behavior, we perform resonant-elastic X-ray scattering (REXS) measurements at the Co K-edge and find evidence for electronic nematicity within the kagomé plane. Resonant scattering provides elemental specificity and is also sensitive to site-specific symmetry through the anisotropic scattering factor \cite{Templeton1980, Templeton1982, Dmitrienko1983}. The latter is enhanced near resonance and arises from the atomic form factor being a tensor, rather than a scalar when the charge distribution around an atomic site breaks spherical symmetry. In non-magnetic materials, this effect can lead to the observation of structurally forbidden peaks, but is not limited to such a case. Importantly, it is uniquely sensitive to additional symmetry breaking, which can lead to an azimuthal dependence of the polarized scattering intensity. This technique has thus been used to identify orbital ordering and nematicity in strongly correlated transition metal systems ranging from manganites \cite{Ishihara1998, Endoh1999, Nakao2017} to cuprates \cite{Achkar2016}. Fig. \ref{Fig3}a shows the temperature-dependent REXS rocking curves of the (003) Bragg peak, which has an order of magnitude enhancement between 200-250 K. Although the in-plane components of the scattering vector are zero, the intensity of this peak still conveys information about the Co site through the total atomic scattering factor (SM). Fig. \ref{Fig3}d shows the energy dependence of the [003] scattering peak in the $\sigma-\sigma'$ and $\sigma-\pi'$ polarization channels, and shows a large peak near E $=7.709$ keV as expected for resonance to the Co atoms. To determine if there is broken symmetry in this temperature range, we measure the scattering intensity in the $\sigma-\sigma'$ polarization channel as a function of sample rotation $\psi$. As shown in Fig. \ref{Fig3}b, the intensity is two-fold symmetric and peaks near $\psi = -45^\circ$. This azimuthal dependence is consistent with the reduction of rotational symmetry depicted in Fig. \ref{Fig3}e, and further analyzed in the SM. Fig. \ref{Fig3}c shows the temperature dependence of the integrated peak intensities in both the $\sigma-\sigma'$ and $\sigma-\pi'$ polarization channels, highlighting the specific region near T$^*$ where the scattering intensity is enhanced. This temperature range is consistent with that in which we observe nematicity through other probes. 

\begin{figure*}[ht!]
 \centering
 \includegraphics[height=0.65\textheight]{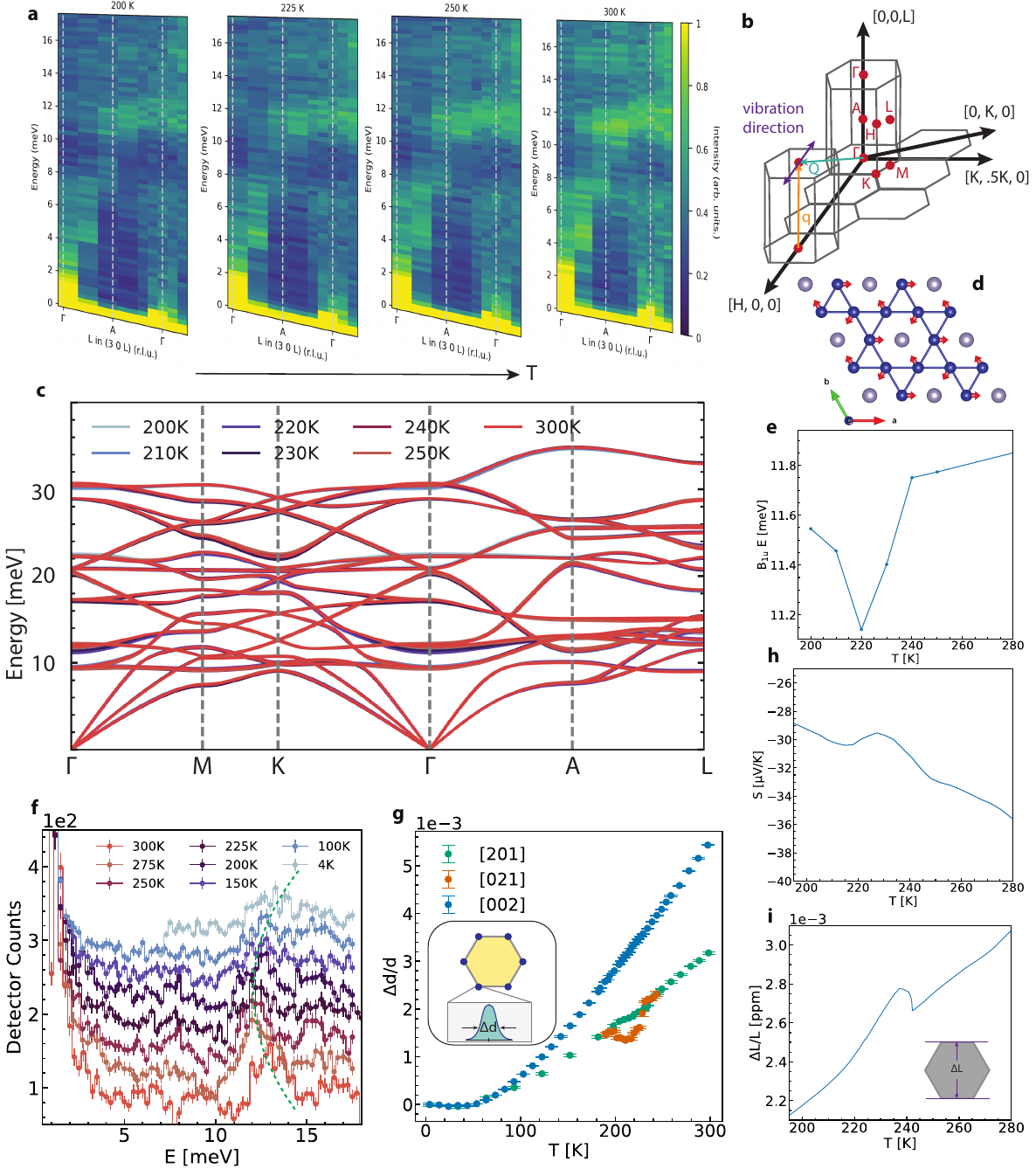}
 \caption{\textbf{Electron-phonon coupling in the nematic regime.} \textbf{(a)} Phonon bandstructure along the $\Gamma$-A-$\Gamma$ direction measured through inelastic neutron scattering (INS) at different temperatures. \textbf{(b)} Depiction of the INS scattering geometry, where $Q$ is the neutron momentum transfer, and $q$ is the phonon wave-vector. Purple arrow indicates the direction of measured atomic vibrations, which are in the kagomé plane. \textbf{(c)} AIMD calculations of the phonon bandstructure for temperatures near the nematicity. \textbf{(d)} Atomic motion of the $B_{1u}$ phonon mode which softens at T$^*$ $=225$ K. \textbf{(e)} $B_{1u}$ phonon energy as a function of temperature, indicating softening through T$^*$. \textbf{(f)} Waterfall plot of energy scans at momentum transfer of $Q$ = [301] from 300 K to 4 K. Error bars are based on Poisson counting statistics and the teal dashed line is a guide to the eye. \textbf{(g)} Neutron Larmor diffraction measurement of the lattice distortion as a function of temperature for three different peaks, [002], [201], and [021]. Inset shows a schematic of measured peak width along the [002] direction. \textbf{(h)} Thermal expansion along the [210] direction measured with a high resolution dilatometer. Inset shows a schematic of the experiment with respect to crystal geometry. \textbf{(i)} Seebeck coefficient measured with a thermal gradient along the $c$-axis.}
 \label{Fig4}
\end{figure*}

\subsection*{Phonon response to nematicity}
To probe the response of the lattice separately from electronic degrees of freedom, we utilize several neutron scattering techniques. First, we use inelastic neutron scattering (INS) to probe the phonon band structure at several different temperatures, as shown in Fig. \ref{Fig4}a. These measurements directly probe phonons which have vibrational direction within the kagomé plane, as depicted in the scattering schematic of Fig. \ref{Fig4}b. INS measurements indicate phonon softening through the nematic temperature range. As shown in Fig. \ref{Fig4}f, the energy of the optical phonon near 13 meV decreases in energy until it reaches a minimum near T$^*\sim225$ K, whereupon it increases in energy as temperature is further decreased. In addition to measuring the phonon spectrum with triple-axis neutron scattering, we also use time-of-flight neutron scattering to uncover dispersionless optical phonon modes associated with geometric frustration (shown in SI Fig. 10) \cite{Yin2020a}. We compare the measured phonon dispersion to finite temperature phonon bandstructure calculations presented in Fig. \ref{Fig4}c, obtained through ab-initio molecular dynamics (AIMD) simulations. These calculations corroborate the softening of an in-plane phonon mode observed in the INS experiments. In particular, AIMD shows that the Co $B_{1u}$ phonon mode, depicted in Fig. \ref{Fig4}d, softens near T$^*$ as shown in Fig. \ref{Fig4}e. 

In addition to INS measurements and AIMD calculations demonstrating phonon softening near T$^*$, neutron Larmor diffraction \cite{Rekveldt_2001, Li2017} further indicates the \pav{presence} of symmetry breaking at this temperature. This technique has been used to study nematic susceptibility in Fe-based superconductors \cite{Wang2018}. As depicted in Fig. \ref{Fig4}d, the distribution of $d$-spacings along a specific crystallographic direction encode the width of the peak into a Larmor precession phase of scattered polarized neutrons. We measured the $d$-spacing distribution along the [002], [201], and [021] directions (Fig. \ref{Fig4}g). Along the [002] direction, there is no anomalous behavior in the temperature region of interest, whereas measurements along the [021] and [201] directions show a distinct change in the $d$-spacing distribution, albeit with different magnitudes. The lack of anomaly along the $c$-axis of the material suggests that the change along the [021] and [201] directions result purely from a distortion within the kagomé plane. The fact that these two Bragg peaks have differently sized distortions further points to the presence of anisotropy through this temperature range. We also directly measure the thermal expansion within the kagomé plane by using a high-resolution dilatometer \cite{Kuchler2023}. As displayed in Fig. \ref{Fig4}h, this measurement further corroborates the anomaly with a sharp change near T $\sim$ 240 K. 
      
Finally, we note that the Seebeck coefficient measured along the $c$-axis shows a distinct anomaly through the nematic regime. Normal resistivity measurements do not show an anomaly at this temperature range (SM), potentially due to the presence of highly dispersive bands close to the Fermi level that dominate the resistivity response. The Seebeck coefficient is related through the Mott formula to the derivative of the electron density of states, $S\propto\frac{\partial n}{\partial E}$, and therefore can be more sensitive than resistivity to changes in electronic structure. Collectively, these probes point towards subtle changes in the lattice which occur simultaneously with the nematicity observed through REXS.

\section*{Discussion}


\begin{figure*}[ht!]
 \centering
 \includegraphics[width=.9\textwidth]{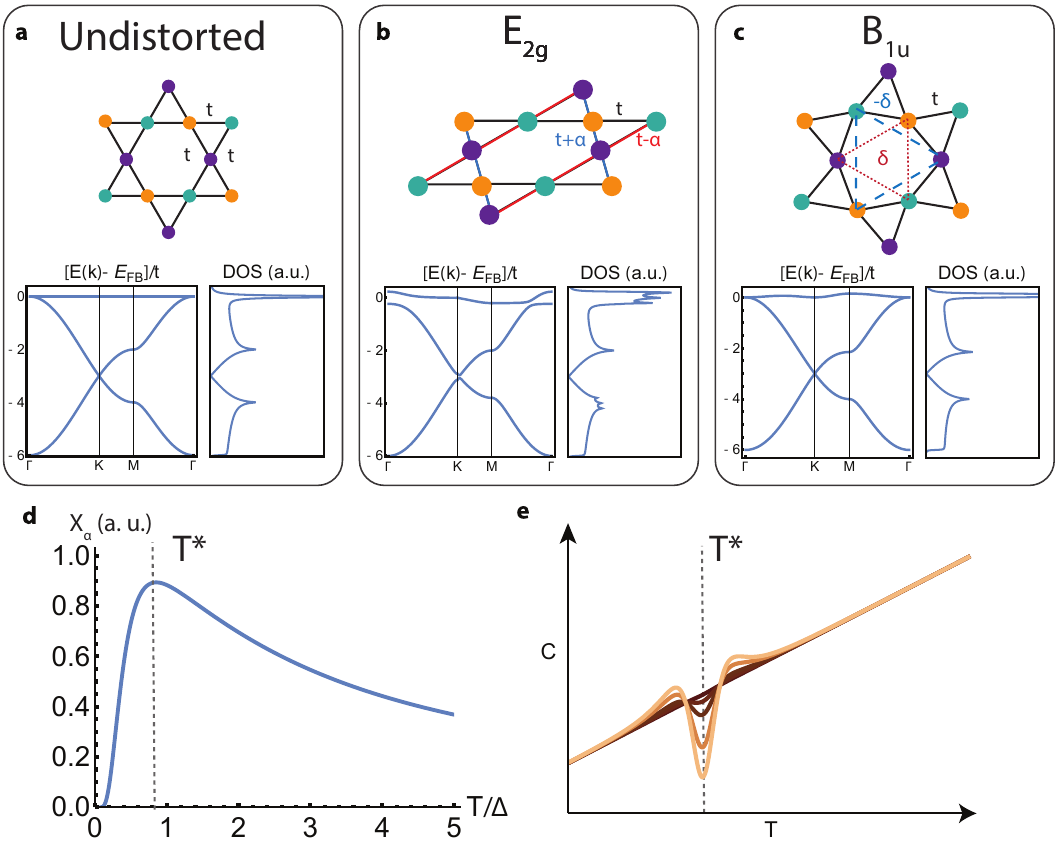}
 \caption{\textbf{Flat band susceptibility to lattice distortions.} \textbf{(a-c)} Illustration of (a) undistorted and (b,c) distorted kagomé plane with corresponding tight-binding model, energy bands and densities of states. (d) Susceptibility of the 2D kagomé lattice tight-binding model to a distortion in (b) as a function of temperature. (e) Qualitative character of specific heat from flat-band-enhanced symmetry lowering, Eq. (1). Regular specific heat is modeled by a linear background. Fainter color corresponds to increasing $\kappa\propto B^2$.}
 \label{Fig5}
\end{figure*}

\subsection*{Effect of flat-bands on nematic susceptibility}
We now demonstrate that all of the observed phenomena can be attributed to thermal activation of kagomé flat band carriers. The flat bands,  characterized by a singular density of states, are highly susceptible to most perturbations. In particular, reduction of $C_6$ symmetry due to strain \pav{or nematicity} causes a splitting and smearing of the density of states peak (Fig. \ref{Fig5}a, see SM), which is linear in perturbation amplitude $\alpha$. \pav{Taking only this effect into account results in thermodynamic susceptibility to  $\alpha$ taking the form $\chi_{FB}(T) \propto\frac{1}{T} \frac{e^{\Delta/T}}{(1+e^{\Delta/T})^2}$} where $\Delta$ is the flat band energy with respect to the Fermi level.  $\chi_{FB}(T)$ has a pronounced maximum at $T^*\approx0.64\Delta$ and accounts for a significant part of susceptibility of the full kagomé lattice model\pav{ $\chi_\alpha(T)$, shown in Fig. \ref{Fig5}d)} (See SM). Based on \pav{$\chi_{FB}(T)$}, we estimate that the distance of the flat-bands to the Fermi level $\Delta \approx30$ meV, which is in excellent agreement with other experimental \pav{estimates} \cite{Meier2020, Huang2020}.


Given the absence of signatures of a phase transition, the symmetry breaking should be attributed to extrinsic factors in form of stress and magnetic field. To take into account both lattice and electronic degrees of freedom, we introduce electronic nematic order $\vec{\Phi}$ that has the same symmetry as non-symmetric strain $\vec{\varepsilon} = (\varepsilon_{xx}-\varepsilon_{yy},2\varepsilon_{xy})$. These are expected to couple linearly to strain and the symmetry-breaking magnetic field bilinears $\vec{\kappa} = (B_x^2-B_y^2, 2 B_x B_y)$, respectively. 
The free energy associated with lowering of the symmetry takes the form 
$$F = \frac{G_\varepsilon}{2} \vec{\varepsilon}^2_{nem}
+\frac{G_{\Phi}(T)}{2} \vec{\Phi}^2
+G \vec{\Phi}\cdot\vec{\varepsilon}_{nem}
-\vec{\sigma}_{nem} \vec{\varepsilon}_{nem} - \vec{\Phi} \vec{\kappa}$$
where \pav{$\vec{\sigma}_{nem}$ is the extrinsic $E_{2g}$ stress and $G_{\Phi}(T) = G_0 - \chi_\alpha(T)$, where the factor $G_0$ describes the effects of interactions and other bands in the spirit of random-phase approximation.} The evidence for stronger electronic effects in the experiments suggests $G_\Phi(T^*)\ll G$, \pav{consistent with a peak of $\chi_\alpha(T)$ at $T^*$,} and the field is small enough such that $G_\varepsilon \kappa \ll G \sigma$. Minimizing this expression as a function of $\vec{\varepsilon}, \vec{\Phi}$, we obtain at $T\approx T^*$
\begin{equation}
F \approx- \frac{1}{2 G_\varepsilon} \frac{G_\Phi(T^*) \vec{\sigma}^2-2G\vec{\kappa}\cdot \vec{\sigma}}{G^2/G_\varepsilon-G_\Phi(T^*)+\frac{|G_\Phi''(T^*)|}{2} \left(T-T^*\right)^2}.
\end{equation}
For $G^2$ close to $G_\varepsilon G_\Phi(T^*)$, this contribution has a strong narrow peak near $T^*$ implying several consequences. First, there is a pronounced peak of strain $|\varepsilon| = -\frac{\partial F}{\partial \sigma}$ at $T^*$. This explains the zero-field thermal-expansion anomalies present in the dilatometry and Larmor-diffraction measurements. Second, specific heat $C = -T\frac{\partial^2 F}{\partial T^2}$ has a 3-peak anomaly near $T^*$, which grows with the strength of the stress (magnetic field). Adding a background linear specific heat (see SM) results in a shape overall consistent with the observed one (Fig. 5 e).  Third, as noted above, the non-symmetric magnetostriction has to be proportional to $\varepsilon_{xy}(T)$. Since it has a pronounced peak at $T^*$, a corresponding peak is expected in non-symmetric magnetostriction. Finally, the softening of phonon frequencies can be explained in this picture too. As shown in Fig. \ref{Fig5}c, the $B_{1u}$ phonon atomic displacements can enhance the effect of next-nearest neighbor hopping, which also distorts the flat bands. \pav{The perturbative phonon} shift \pav{contains a contribution} (see SM) $\delta \omega_0 \propto- V^2 \chi_{B_{1u}}(T)$, which is also expected to have a prominent peak at a finite temperature, \pav{although in the model it does not necessarily coincide with $T^*$.}
 
Within the context of other kagomé metals, our observations are unique in so much as they show nematicity without the presence of other symmetry breaking phases such as magnetism, superconductivity, or charge-density-wave (CDW) order, and because of its transience at relatively high temperatures above 200 K. In the V-based ``135" compounds, the nematicity has been associated with the C$_2$ symmetry breaking provided by the CDW phase \cite{Nie2022, Xu2022, Li2022a, Li2023b}. Recent studies of the Ti-based ``135" compounds have shown evidence for electronic nematicity without the CDW ordering \cite{Li2023, Jiang2023}, although it is not clear what the origin of the electronic nematicity is, as the Fermi level seems to be away from the van Hove filling, and there is not a clear thermodynamic signature. 

In contrast to the other kagomé metals, CoSn has the flat bands very close to the Fermi level, which we suggest drives the enhanced nematic correlations, albeit without a clear transition. Note that in hexagonal systems, nematic transition belongs to the three-state Potts universality class \cite{Fernandes2020} and should be generically first order \pav{that would make signatures of a true transition stronger}. Thus our results point to CoSn being right at the brink of a nematic instability at finite temperature. Tuning the system with hydrostatic or c-axis pressure or doping may therefore reveal new thermodynamic phases and make fluctuations even stronger. We note that the concomitant enhancement of thermoelectric coefficient further suggests that such critical kagomé nematics may have superior thermoelectric properties.

More broadly, our results highlight how thermal fluctuations can be beneficial for engineering strongly correlated phases. Despite the relatively high temperatures in our experiments, the effect we observe can be still attributed to a competition between quantum and classical effects. In particular, interference of electronic waves on the kagomé lattice tends to localize the electrons, while classical fluctuations of the lattice disrupt this geometric effect. Contrary to the Mott-Hubbard paradigm in which the potential energy $U$ arises from classical electron-electron interactions, in this case, it is the quantum effects that favor electron localization. The close competition between the two, however, results in anomalous softening of the lattice and pronounced magnetoentropic responses. Intriguingly, these competing tendencies are already evident in the dilute, non-degenerate limit for the flat band \pav{carriers}. This opens future perspectives for observation of exotic interaction-driven behavior in non-degenerate electron systems with flat bands, relaxing the requirement for the flat band to be at the Fermi level. Having a flat band away from the Fermi-level in gapped systems also allows a fresh view on semiconductor physics, where the interaction effects due to flat bands endow the quasiparticles with large amounts of entropy, which may lead to realization of favorable thermoelectric properties \cite{Dresselhaus2007} or other technologically useful phenomena.

\section*{Methods}

\subsection*{Single crystal growth}
We synthesized high-quality single crystals of CoSn through the Sn self-flux method. A mixture of Co chunks and Sn pellets were weighed in a molar ratio of 1:4 into a crucible. The mixture-filled crucible was flame-sealed in an evacuated quartz tube and was subsequently heated up to 400$^\circ$C from room temperature at a rate of 100$^\circ$C/h, then dwelled at 400$^\circ$C for 2 hours. Next, the materials were heated to 950$^\circ$C at 100$^\circ$C/h and were then held at 950$^\circ$C for 10 hours. The samples were then cooled to 650$^\circ$C at a rate of 3$^\circ$C/h, and were subsequently. This was followed by several days of annealing at this temperature after which centrifugation was performed to remove the excess flux. The resulting products of CoSn single crystals are hexagonal prisms approximately half-centimeter long and have a metallic luster with lattice constants $a = b = 5.272\ \si{\angstrom}$ and $c = 4.246\ \si{\angstrom}$ as measured with single crystal neutron diffraction. 

\subsection*{Heat capacity measurements}
Heat capacity measurements were taken with the Heat Capacity module of the Quantum Design PPMS Dynacool. The sample was mounted onto the stage with N-grease and the heat capacity was measured with the relaxation time method in addition to subtracting the addenda of the grease heat capacity. At each temperature, the heat capacity was measured three times and averaged. To achieve different magnetic field directions, the sample was mounted in different orientations such that the field was applied along the [210] and [001] directions. The heat capacity measured for applied fields of 0 T and 9 T in the kagomé plane are displayed in Supplementary Figure 3 on the left. On the right is shown the first derivative of heat capacity at zero field with respect to temperature, which is fit to a model of electronic, phononic, and anomalous contributions in the points and solid line, respectively. 

\subsection*{Thermoelectric transport}
Electrical and thermal transport were measured with the ETO and TTO modules of the Quantum Design PPMS Dynacool system. Device contacts were made with silver epoxy H20E. For the thermal conductivity measurement, thermal gradient was applied along the c-axis of the material and thermometers were also placed along the c-axis. For the electrical transport measurements, the applied current along is the c-axis, voltage measured along the c-axis. 

\subsection*{Dilatometry}
Magnetostiction and thermal expansion experiments were performed with a commercially available mini-dilatometer \cite{Kuchler2023} which is compatible with the sample environment provided by the Quantum Design PPMS Dynacool. This device can be rotated relative to the applied magnetic field to find the magnetostriction for field applied along specific crystallographic axes. The sample was mounted such that the change in length was measured along the $[210]$ direction for fields applied along the $[110]$ and $[010]$ directions.

\subsection*{Neutron Larmor diffraction}
High resolution neutron Larmor diffraction was used to measure the lattice spacings along different crystallographic directions\cite{Rekveldt_2001, Wang2018} at the HB-1 beamline at the High Flux Isotope Reactor (HFIR) at Oak Ridge National Laboratory \cite{Li2017}.

\subsection*{Resonant elastic X-ray Scattering (REXS)}
Resonant elastic X-ray scattering measurements were carried out at the 4-ID beamline at NSLS-II in Brookhaven National Laboratory. Single crystals were cleaved and mounted into the [H0L] scattering plane orientation with GE varnish and aligned to the [203] and [003] peaks. Temperature control was attained with a diplex closed cycle cryostat. The photon energy was selected to be at the Co K-edge, 7.709 keV. Energy scans with fixed Q from 7.6 keV to 7.8 keV at the [003] peak demonstrate the Co K-edge resonance. Subsequent measurements were carried out at either 7.709 keV or 7.696 keV. Incoming polarization was fixed in the $\sigma$ orientation, while outgoing polarization was detected in the $\sigma$' or $\pi$' channels via a graphite (006) crystal analyzer.

\subsection*{Single crystal neutron diffraction}
Single crystal neutron diffraction experiments were done at the BL-9 CORELLI instrument at the Spallation Neutron Source (SNS) of Oak Ridge National Lab. A single crystal of 80 mg was loaded into the instrument with two different orientations, one with the (HK0) scattering orientation and the other along the (H0L) scattering orientation. The samples were cooled in zero field to temperatures of 150 K, 220 K, and 300 K. At each of these temperatures, a magnetic field of 4.75 T was applied perpendicular to the c-axis and parallel to the c-axis for the H0L and HK0 orientations, respectively. 

\subsection*{Inelastic neutron scattering (INS)}
Inelastic neutron scattering measurements were performed at Oak Ridge National Laboratory (ORNL). Specifically, the triple-axis measurements were carried out at the HB-3 beamline at the High Flux Isotope Reactor (HFIR) and the time-of-flight measurements were carried out at the ARCS beamline at the Spallation Neutron Source (SNS). HFIR experiments were carried out by selecting a final scattered neutron energy of $E_f = 14.7$ meV and varying the incident neutron energy in constant Q mode. Horizontal collimation settings of 48’-40’-sample-40’-120’ were used. ARCS measurements were done with incident neutron energy of $E_i = 35$ meV. 

\subsection*{Ab-initio molecular dynamics calculations}
To capture the renormalized phonon dispersions at finite temperature, the temperature-dependent interatomic force constants (IFCs) were extracted by combining ab initio molecular dynamics (AIMD) simulations and the temperature-dependent effective potential (TDEP) technique \cite{hellman2013temperature,hellman2011lattice}. The AIMD simulations were performed within the density functional theory (DFT) framework implemented in the Vienna Ab-initio Simulation Package (VASP) \cite{kresse1993ab,kresse1996efficiency,kresse1996efficient}. The simulations used the projector-augmented wave formalism \cite{blochl1994projector} with exchange-correlation energy functional parameterized by Perdew, Burke, and Ernzerhof within the generalized gradient approximation \cite{perdew1996generalized}. Before AIMD simulations, the crystal structure was fully relaxed with energy and Hellmann–Feynman force convergence thresholds of $10^{-6}$ eV and $10^{-4}$ eV \r{A}, respectively, and the difference between the optimized lattice constants and the experimental values is within 1\%.  The AIMD simulations were performed on a 3×3×3 supercell, 162 atoms in total. The electronic self-consistent loop convergence was set to $10^{-5}$ eV. A single $\Gamma$-point k-mesh with a plane-wave cut-off energy of 350 eV was used to fit the effective energy surface using the TDEP method. The simulations were performed at 200, 210, 220, 230, 240, 250, and 300 K, with the NVT ensemble using a Nose–Hoover thermostat. All the simulations were run for 10 ps with a timestep of 2 fs, and the initial 1 ps’s information was discarded due to the nonequilibrium.

\subsection*{Analytical theory}

\pav{Susceptibility, band structures and densities of states in Fig. 5 were calculated using a single-orbital tight-binding model on a kagomé lattice, including effects of strain \cite{liu2020_th} or $B_{1u}$ distortion via change in hopping integrals, consistent with symmetry of the distortion, as shown in Fig. 5 (b),(c) (see SM for details).}

\section*{Acknowledgements}
NCD and MM acknowledge support from the US Department of Energy (DOE), Office of Science (SC), Basic Energy Sciences (BES), Award No. DE-SC0020148. TN is supported by NSF Designing Materials to Revolutionize and Engineer our Future (DMREF) Program with Award No. DMR-2118448. PS acknowledges the support of DOE Award No. DE-SC0021940. TB acknowledges support from NSF Convergence Accelerator Award No. 2235945. ML acknowledges the support from the Class of 1947 Career Development Chair and support from R Wachnik. The research on neutron scattering used resources at Oak Ridge National Laboratory's High Flux Isotope Reactor (HFIR) and Spallation Neutron Source (SNS) which are sponsored by the U.S. Department of Energy, Office of Basic Energy Sciences. The X-ray scattering measurements used resources of the National Synchrotron Light Source II, a U.S. Department of Energy (DOE) Office of Science User Facility operated for the DOE Office of Science by Brookhaven National Laboratory under Contract No. DE-SC0012704. \\

\section*{Author Contributions}
 NCD and ML conceived and ML supervised the project. NCD and TN performed the transport measurements with the support from ML. NCD and MM synthesized the materials. NCD, TN, MM, AB and PS performed neutron scattering measurements with help from TJW, SC, DLA, FL, FY, KB, FF, and MM. NCD, and MM performed magnetometry. NCD, MM, RO, PS, and AB, performed X-ray scattering measurements with help from CSN. NCD performed the dilatometry experiments. PV developed the theory. YQ performed the ab initio calculations with support from BL. NCD, PV and ML wrote the paper with input from all authors.\\

\section*{Competing Interests}
The authors declare no competing financial interests.\\


\end{document}


\begin{center}
\large
\textbf{Supplementary Information\\ Incipient nematicity from electron flat-bands in a kagomé metal}

\vspace{0.5cm}

\normalsize
Nathan C. Drucker$^{1, 2, \dagger}$, Thanh Nguyen$^{1, 3}$, Manasi Mandal$^{1, 3}$, Yujie Quan $^{4}$, Artittaya Boonkird $^{1,3}$, Phum Siriviboon$^{1, 5}$, Ryotaro Okabe$^{1, 6}$, Fankgang Li$^{7}$, Kaleb Burrage$^{7}$, Fumiaki Funuma$^{7}$, Masaaki Matsuda$^{7}$, Douglas L. Abernathy$^{7}$, Travis Williams$^{7}$, Songxue Chi$^{7}$, Feng Ye$^{7}$ Christie Nelson$^{8}$, Bolin Liao$^{9,10}$, Pavel Volkov$^{9,10,\dagger}$, and Mingda Li$^{1, 3, \dagger}$\\
\vspace{3mm}
\small
\textit{$^{1}$Quantum Measurement Group, MIT, Cambridge, MA 02138, USA\\
$^{2}$John A. Paulson School of Engineering and Applied Sciences, Harvard University, Cambridge, MA 02138, USA\\
$^{3}$Department of Nuclear Science and Engineering, MIT, Cambridge, MA 02139, USA\\
$^{4}$Department of Mechanical Engineering, University of California, Santa Barbara, Santa Barbara, CA 93106, USA\\
$^{5}$Department of Physics, MIT, Cambridge, MA 02139, USA\\
$^{6}$Department of Chemistry, MIT, Cambridge, MA 02139, USA\\
$^{7}$Neutron Scattering Division, Oak Ridge National Laboratory, Oak Ridge, TN 37831, USA\\
$^{8}$National Synchrotron Light Source II, Brookhaven National Laboratory, Upton, NY 11973, USA\\
$^{9}$Department of Physics, Harvard University, Cambridge, MA 02138, USA\\
$^{10}$Department of Physics, University of Connecticut, Storrs, CT 06269, USA}\\
\vspace{3mm}
$^{\dagger}$Corresponding authors. 

\end{center}
\thispagestyle{empty}

\clearpage

\normalsize
\section*{Supplementary Note 1: Samples}\label{sec:Single crystal growth}
We synthesized high-quality single crystals of CoSn through the Sn self-flux method. A mixture of Co chunks and Sn pellets were weighed in a molar ratio of 1:4 into a crucible. The mixture-filled crucible was flame-sealed in an evacuated quartz tube and was subsequently heated up to 400$^\circ$C from room temperature at a rate of 100$^\circ$C/h, then dwelled at 400$^\circ$C for 2 hours. Next, the materials were heated to 950$^\circ$C at 100$^\circ$C/h and were then held at 950$^\circ$C for 10 hours. The samples were then cooled to 650$^\circ$C at a rate of 3$^\circ$C/h, and were subsequently. This was followed by several days of annealing at this temperature after which centrifugation was performed to remove the excess flux. 

The resulting products of CoSn single crystals are hexagonal prisms approximately half-centimeter long and have a metallic luster with lattice constants $a = b = 5.272$ \r{A} and $c = 4.246$ \r{A} as measured with single crystal neutron diffraction. 

\begin{figure}[ht!]
 \centering
 \includegraphics[width=0.8\columnwidth]{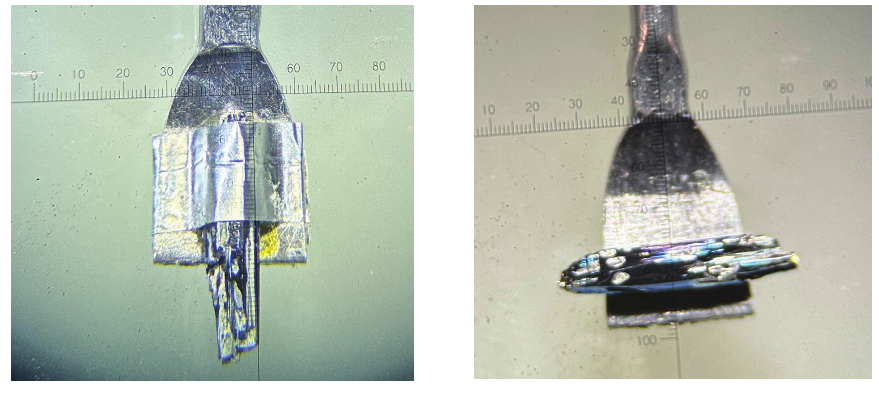}
 \caption{\textbf{Mounted CoSn single crystal for neutron diffraction.} Image of CoSn single crystal mounted on an aluminum plate in preparation for elastic neutron diffraction. The lateral size of the crystal is $\sim6$ mm$\times$2 mm.}
 \label{supfig:sampleimage-1}
\end{figure}

\section*{Supplementary Note 2: Electrical and thermal 
transport}\label{sec:electric-transport}

Electrical and thermal transport were measured with the ETO and TTO modules of a Quantum Design PPMS Dynacool system. Device contacts were made with silver epoxy H20E. For the thermal conductivity measurement, thermal gradient was applied along the c-axis of the material and thermometers were also placed along the c-axis. For the electrical transport measurements, the applied current along is the c-axis, voltage measured along the c-axis. The electrical and thermal transport are displayed in Supplementary Figure 2 on the left and right, respectively.

\section*{Supplementary Note 3: Heat capacity }\label{sec:heat capacity}

Heat capacity measurements were taken with the Heat Capacity module of the Quantum Design PPMS Dynacool. The sample was mounted onto the stage with N-grease and the heat capacity was measured with the relaxation time method in addition to subtracting the addenda of the grease heat capacity. At each Temperature, the heat capacity was measured three times and averaged. To achieve different magnetic field directions, the sample was mounted in different orientations such that the field was applied along the [210] and [001] directions. The heat capacity measured for applied fields of 0 T and 9 T in the kagomé plane are displayed in Supplementary Figure 3 on the left. On the right is shown the first derivative of heat capacity at zero field with respect to temperature, which is fit to a model of electronic, phononic, and anomalous contributions in the points and solid line, respectively.

\begin{figure}[ht!]
 \centering
 \includegraphics[width=.85\columnwidth]{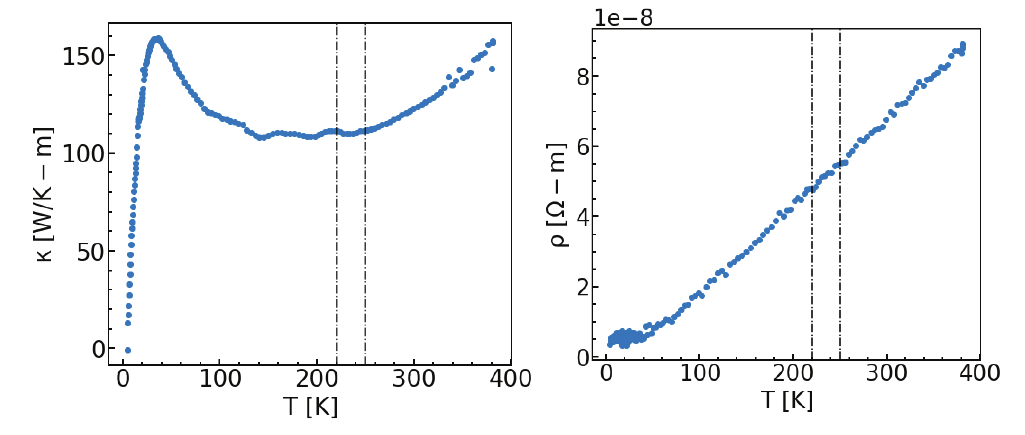}
 \caption{\textbf{CoSn transport} Left, thermal conductivity along the c-axis for thermal gradien along the c-axis measured from 2K to 300K. Right, Electrical resistivity for current and voltage direction along the c-axis, measured for temperatures from T=2K to T=300K. In each plot, dashed lines indicate boundary of anomalous behavior found in the thermoelectric measurement presented in the main text Figure 4.}
 \label{supfig:transport}
\end{figure}

\begin{figure}[ht!]
 \centering
 \includegraphics[width=.85\columnwidth]{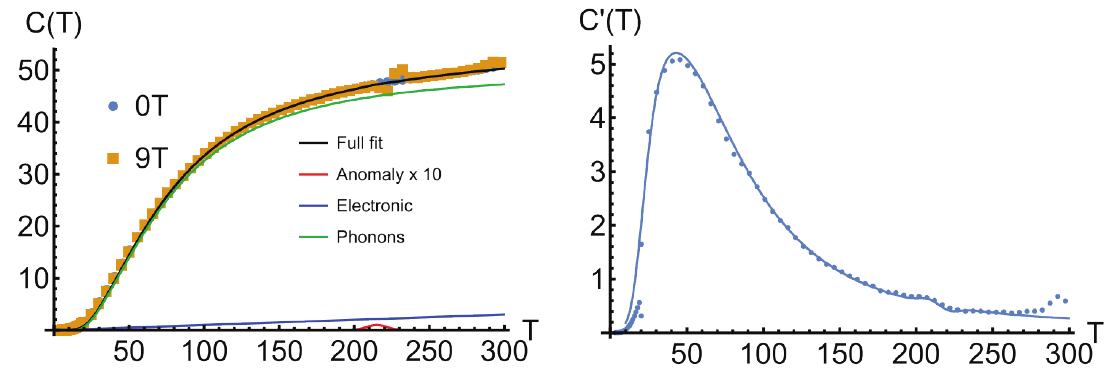}
 \caption{\textbf{Heat capacity anomaly in zero field.} Left, points represent experimentally measured heat capacity from T=2K to T=300K in fields applied within the kagomé plane B=0T and B=9T. Solid lines indicate fits using a basic model phonon and electronic contributions to specific heat, with an additional small anomaly. Right, points are the numerical derivative of zero-field heat capacity data, and the solid line is the derivative of the model. There is a small kink in the derivative of the zero-field heat capacity near T=225K }
 \label{supfig:heat capacity}
\end{figure}

\section*{Supplementary Note 4: REXS measurements}\label{sec:rexs}

X-ray scattering measurements were carried out at the 4-ID beamline at NSLS-II. Single crystals were cleaved and mounted into the [H0L] scattering plane orientation with GE varnish and aligned to the [203] and [003] peaks. Temperature control was attained with a diplex closed cycle cryostat. The photon energy was selected to be at the Co K-edge, 7.709 keV. Energy scans with fixed Q from 7.6 keV to 7.8 keV at the [003] peak demonstrate the Co K-edge resonance. Subsequent measurements were carried out at either 7.709 keV or 7.696 keV. Incoming polarization was fixed in the $\sigma$ orientation, while outgoing polarization was detected in the $\sigma$' or $\pi$' channels via a Graphite (006) crystal analyzer. The scattering collected at the [003] peak in the $\sigma - \pi$' channel also shows a significant change with temperature through the anomaly at $T=225$ K.

\begin{figure}[ht!]
 \centering
 \includegraphics[width=1.0\columnwidth]{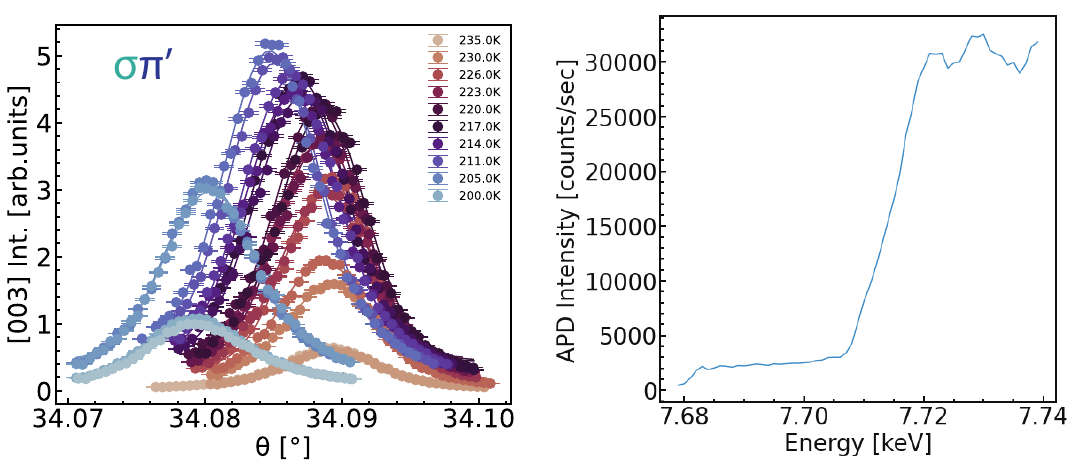}
 \caption{\textbf{REXS at the Co K-edge} Left, rocking curves taken at temperatures near the anomaly for X-rays in the polarization-flip channel $\sigma\pi'$. Right, fluorescence scan near the Co K-edge measured at room temperature.}
 \label{supfig:REXS data}
\end{figure}

\begin{figure}[ht!]
 \centering
 \includegraphics[width=.8\columnwidth]{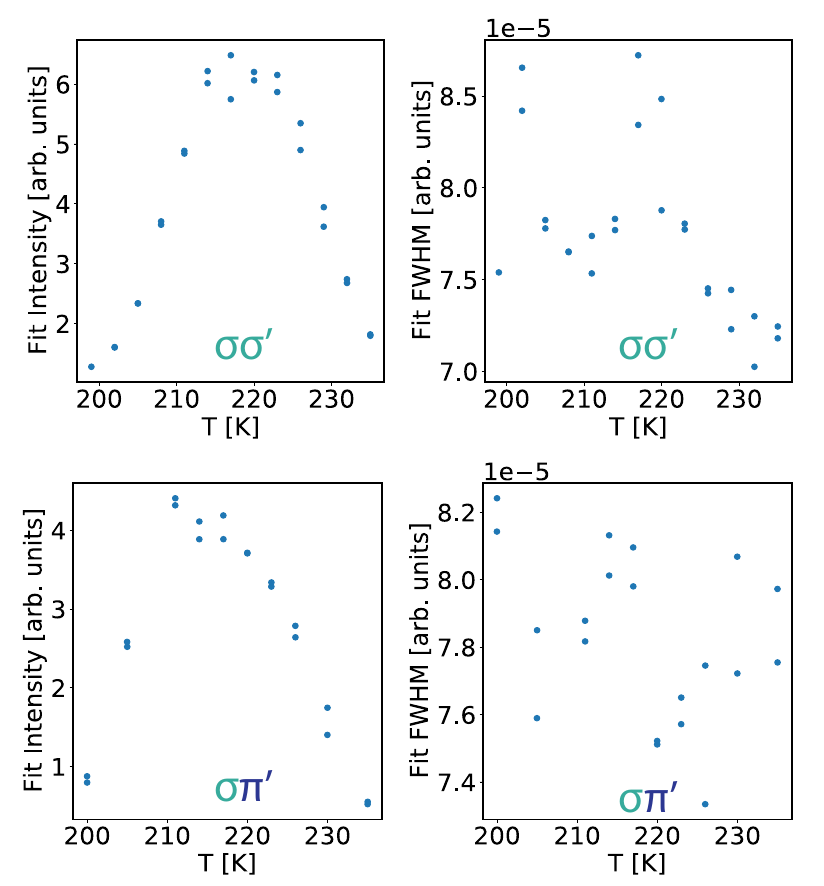}
 \caption{\textbf{REXS intensity and width through the anomaly} [003] Peak maximum intensity (left) and full-width-half-max (right) based on Gaussian fits of the rocking curves taken in both polarization channels (top, bottom respectively). }
 \label{supfig:REXS FWHM}
\end{figure}

\section*{Supplementary Note 5: Anisotropic X-ray Structure Factor}\label{sec:axs}
\begin{figure}[ht!]
 \centering
 \includegraphics[width=.8\columnwidth]{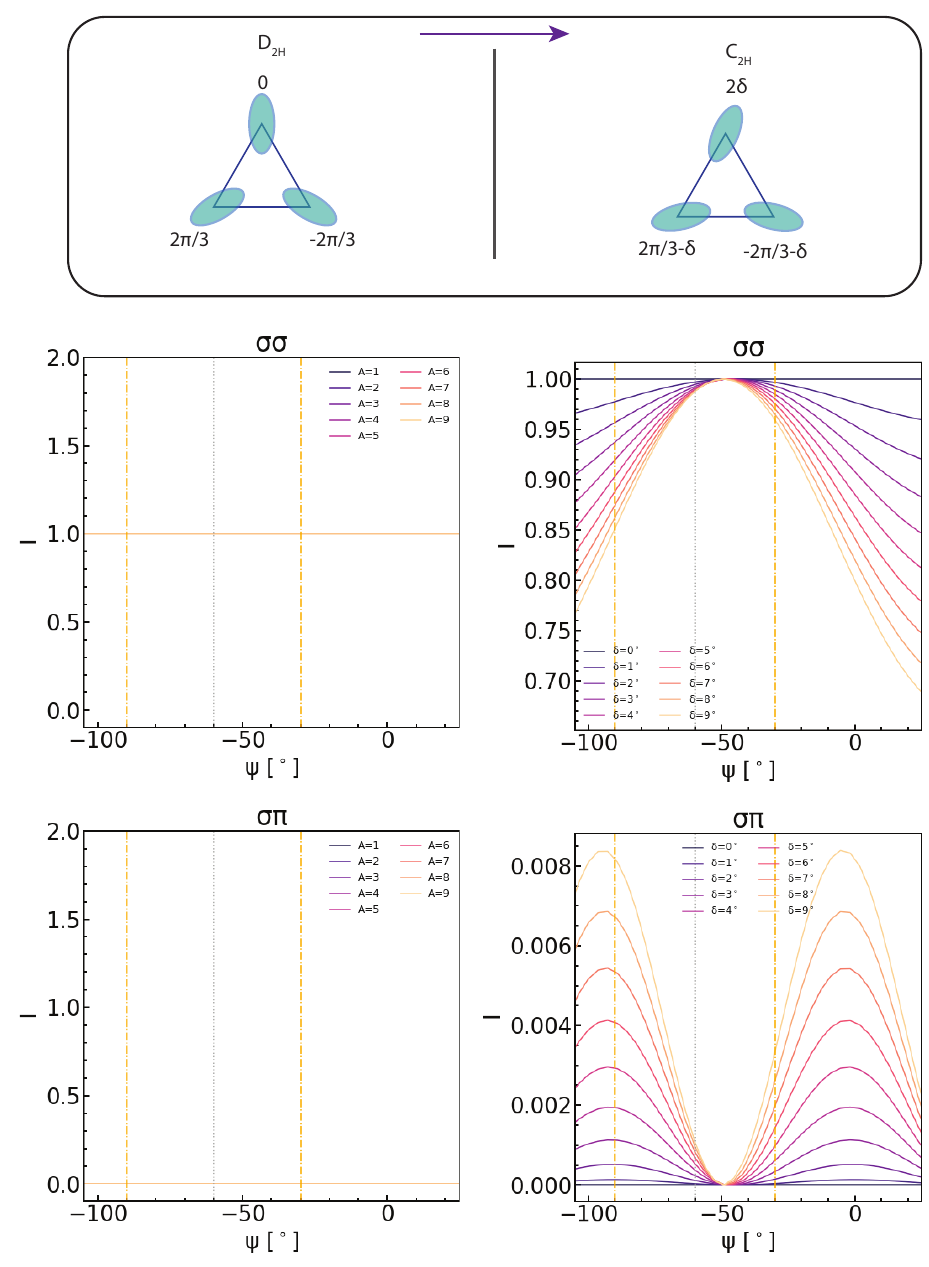}
 \caption{\textbf{Modeling anomalous X-ray scattering} Azimuthal dependence of the REXS intensity in different polarization channels for an undistorted (left) and nematic (right) charge distribution. Note that even when the charge distribution is anisotropic but there is no nematicity, there is no azimuthal dependence on the scattering intensity.   }
 \label{supfig:REXS sims}
\end{figure}

When incident X-rays are resonant to an atomic transition, it becomes necessary to include higher order terms in the interaction between the photons and electrons in the sample. As a result, the so-called anomalous X-ray scattering factor (AXS) introduces additional terms on top of the standard Thompson scattering $f_0$ which arises from dipole-dipole scattering $f(E) = f_0 + f'(E) + if''(E)$ \cite{Dmitrienko1983, Templeton1980}. These terms have quadrupole-quadrupole and dipole-quadrupole contributions. In addition to the AXS enhancement, scattering from anisotropic charge distributions can further modify the scattering cross-section, leading to phenomena including but not limited to the observation of forbidden Bragg peaks near resonance (Templeton scattering). The consequence of the anisotropic charge distribution is that the scattering factor $F(E, Q)$ becomes a tensor, rather than a scalar. This tensor must respect the underlying symmetry of the unit cell, and is thus sensitive to symmetry breaking. Such symmetry breaking can be provided by magnetism, orbital order, or nematicity. In addition to the anomalous scattering factor being energy dependent, it also depends on the azimuthal angle $\psi$ between the incident X-rays and the motif that they scatter from. In sum:

$$F(E, Q, \psi) = \sum_n \varepsilon^i U(\psi)V_n F_n (E,Q) V_n^TU^T(\psi)\varepsilon^f$$

Where E is the incident photon energy, Q is the scattering momentum, n are the atoms in the unit cell, $\psi$ is the relative angle between the lab frame and the crystallographic basis, and $\varepsilon^{i,f}$ are the incoming and outgoing X-ray polarizations. For the $\sigma \sigma'$ channel $\varepsilon^i = \varepsilon^f = (1, 0, 0)$, whereas for the $\sigma \pi'$ channel $\varepsilon^i = (1,0, 0)$ and $\varepsilon^f = (0, 0, 1)$ The U and V matrices are rotations between the lab-frame and crystal coordinates, and orbital motif and crystal coordinates, respectively. 

Because of the anomalous X-ray effect, $F_n$ is a diagonal tensor in the basis of the (orbital) motif:

$$ F_n(E,Q) = \begin{pmatrix}
F_{xx} & 0 & 0\\
0 & F_{yy} & 0\\
0 & 0 & F_{zz}
\end{pmatrix}$$

In the isotropic case within the xy plane, $F_{xx} = F_{yy}$. In the anisotropic case (within the a-b plane), we can parameterize the in-plane anisotropy as a factor $a=F_{xx}/F_{yy}$. 

$$ F_n(E,Q) = \begin{pmatrix}
a & 0 & 0\\
0 & 1 & 0\\
0 & 0 & c
\end{pmatrix}$$

To construct the scattering intensity as a function of asymmetry $a$ at the [003] scattering peak, we simply use the formula for total structure factor from the Co atoms. We note that even though the scattering vector $Q$ does not have an in-plane component, the total Intensity still depends on the in-plane structure factors due to the phase factor. 

$$I(Q) = \left|\sum_{n}F_n' (Q) e^{iQ\cdot R_n}\right|^2$$

Where $Q=[003]$ and $R_n$ are the three sites of the Co atoms in the unit cell and $F_n'(Q)$ is the structure factor after the proper rotations defined above. An example of this motif that is anisotropic at the Co site is shown in SI Figure 6. While the anisotropy changes the total magnitude of the scattering factor, note that it alone does not break the rotational symmetry of the unit cell, and thus does not induce an azimuthal dependence in either the $\sigma \sigma'$ or the $\sigma \pi'$ polarization channels. We can model the rotational symmetry breaking by introducing an angular perturbation to the anisotropic Co sites, which introduces an azimuthal dependence of both polarization channels.   

This azimuthal dependence is explicitly shown for the two cases in SI figure 6, based on numerical calculations of the total Intensity I(Q) defined above. This dependence can also be worked out analytically for arbitrary perturbations $\delta_i$ and anisotropy factor $a$ for each Co site $i$ in the unit cell. It can be shown that 

$$\delta I_{\sigma \sigma'} \propto [2\delta_1 - \delta_2 - \delta_3](1-a)\sin2\psi + [\delta_2 - \delta_1]\sqrt{3}(a+1)\cos2\psi$$
$$\delta I_{\sigma \pi'} \propto (1-a)^2[(2\delta_1 - \delta_2 - \delta_3)\cos2\psi +[\delta_2 - \delta_3]\sqrt{3}\sin2\psi]^2$$

The $\sigma \sigma'$ polarization channel is two-fold symmetric whereas the $\sigma \pi'$ is four-fold symmetric. For small $\delta_i$ the $\sin2\psi$ term in $\sigma \sigma'$ dominates and the azimuthal dependence is peaked near $\psi = \pi/4$. These analytical results are confirmed through the numerical calculations shown in SI Figure 6, and also are corroborated by the data in the main text Figure 3, where the azimuthal intensity in the $\sigma \sigma'$ channel is two-fold and peaked near $\psi = \pi/4$. 

Experimentally, the azimuthal scans were performed by rotating the sample azimuth, and then optimizing the peak position at each azimuthal angle to correct for possible translations of the beam relative to the the sample and additional tilting effects.

\section*{Supplementary Note 5: Single Crystal Neutron Diffraction}\label{sec:neutron-diffraction}

Single crystal neutron diffraction experiments were done at the BL-9 CORELLI instrument at the Spallation Neutron Source (SNS) of Oak Ridge National Lab. A single crystal of 80 mg was loaded into the instrument with two different orientations, one with the (HK0) scattering orientation and the other along the (H0L) scattering orientation. The samples were cooled in zero field to temperatures of 150 K, 220 K, and 300 K. At each of these temperatures, a magnetic field of 4.75 T was applied perpendicular to the c-axis and parallel to the c-axis for the H0L and HK0 orientations, respectively. 

The diffraction patterns for the (HK0) and (H0L) planes are shown in Supplementary Figures 7 and 8, respectively. To within the resolution of the instrument of $\Delta Q/ Q \sim 1e-3$, there is not an apparent change in the structure as temperature is cooled or field is changed. In neither the (H0L) plane nor the (HK0) planes are there peaks other than the structural Bragg peaks. This lack of new peaks around the Bragg positions agrees with the hypothesis that the anomalies are not due to emergent magnetic or charge density wave orderings. 

\begin{figure}[ht!]
 \centering
 \includegraphics[width=1.0\columnwidth]{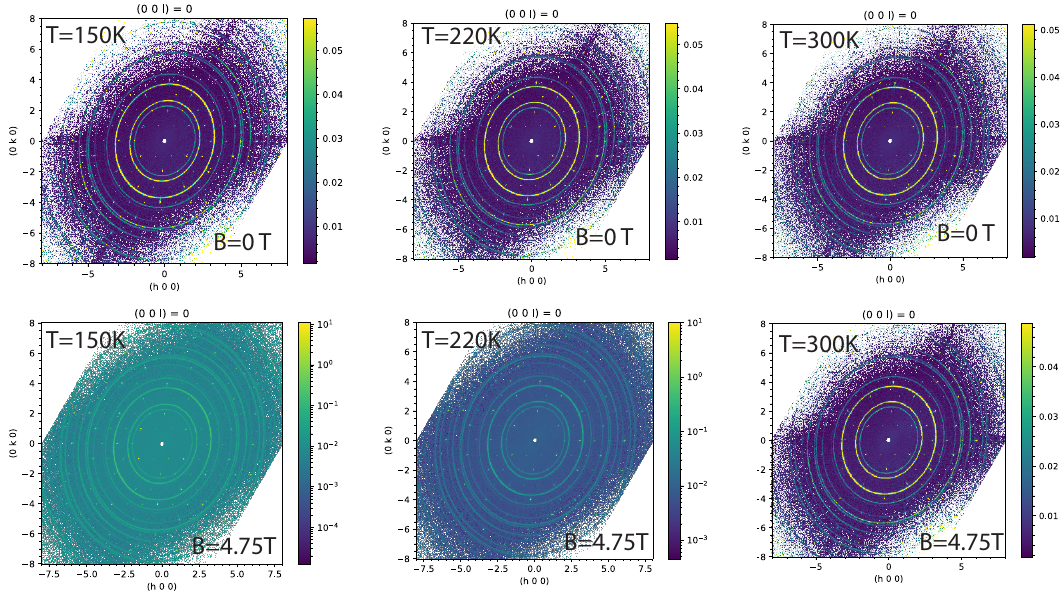}
 \caption{\textbf{Neutron diffraction in the (HK0) plane.} Time of flight neutron diffraction patterns at temperatures of 150 K, 220 K, 300 K left, right, middle) and magnetic field of 4.75 T (top, bottom, respectively). Within the (HK0) scattering plane, there are only structural Bragg peaks resolved, confirming the lack of space group change, magnetic order, or CDW order.}
 \label{supfig:HK0 diff}
\end{figure}

\begin{figure}[ht!]
 \centering
 \includegraphics[width=1.0\columnwidth]{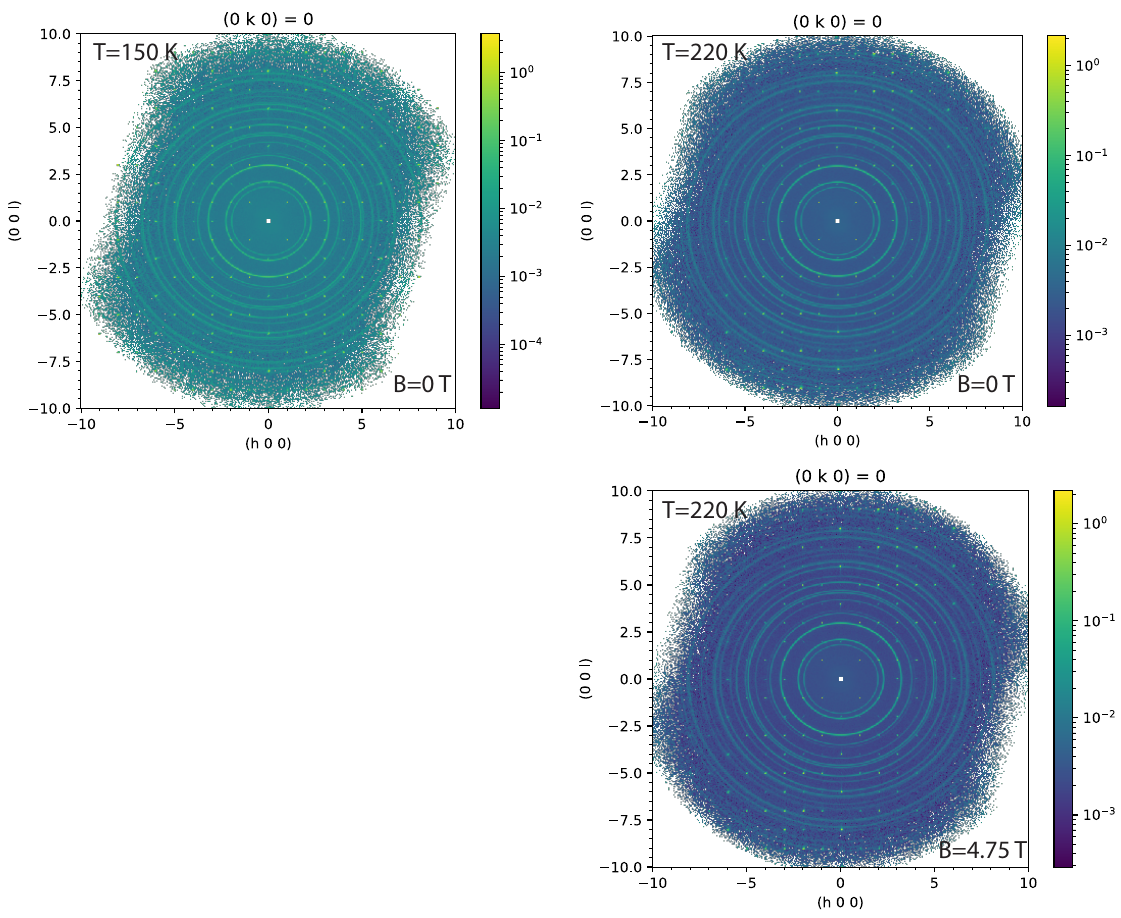}
 \caption{\textbf{Neutron diffraction in the (H0L) plane.} Time of flight neutron diffraction patterns at temperatures of 150 K and 220 K (left, right respectively). Within the (H0L) scattering plane, there are only structural Bragg peaks resolved, confirming the lack of space group change, magnetic order, or CDW order.}
 \label{supfig:HL0 diff}
\end{figure}

\section*{Supplementary Note 6: Ab-initio molecular dynamics calculations }\label{AIMD}
To capture the renormalized phonon dispersions at finite temperature, the temperature-dependent interatomic force constants (IFCs) were extracted by combining ab initio molecular dynamics (AIMD) simulations and the temperature-dependent effective potential (TDEP) technique \cite{hellman2013temperature,hellman2011lattice}. 
The AIMD simulations were performed within the density functional theory (DFT) framework implemented in the Vienna Ab initio Simulation Package (VASP) \cite{kresse1993ab,kresse1996efficiency,kresse1996efficient}. The simulations used the projector-augmented wave formalism \cite{blochl1994projector} with exchange-correlation energy functional parameterized by Perdew, Burke, and Ernzerhof within the generalized gradient approximation \cite{perdew1996generalized}. 
Before AIMD simulations, the crystal structure was fully relaxed with energy and Hellmann–Feynman force convergence thresholds of $10^{-6}$ eV and $10^{-4}$ eV/\AA, 
respectively, and the difference between the optimized lattice constants and the experimental values is within 1\%.  
The AIMD simulations were performed on a 3×3×3 supercell, 162 atoms in total. The electronic self-consistent loop convergence was set to $10^{-5}$ eV. A single $\Gamma$-point k-mesh with a plane-wave cut-off energy of 350 eV was used to fit the effective energy surface using the TDEP method. The simulations were performed at 200, 210, 220, 230, 240, 250, and 300 K, with the NVT ensemble using a Nose–Hoover thermostat. 
All the simulations were run for 10 ps with a timestep of 2 fs, and the initial 1 ps’s information was discarded due to the nonequilibrium. 


\section*{Supplementary Note 7: Inelastic Neutron Scattering}\label{sec:INS}

Inelastic neutron scattering measurements were performed at Oak Ridge National Laboratory (ORNL). Specifically the triple-axis measurements were carried out at the HB3 beamline at the High Flux Isotope Reactor (HFIR) and the time-of-flight measurements were carried out at the ARCS beamline at the Spallation Neutron Source (SNS). HFIR experiments were carried out by selecting a final scattered neutron energy of $E_f = 14.7$ meV and varying the incident neutron energy in constant Q mode. Horizontal collimation settings of 48’-40’-Sample-40’-120’ were used. ARCS measurements were done with incident neutron energy of $E_i = 35$ meV.

\begin{figure}[ht!]
 \includegraphics[width=.8\columnwidth]{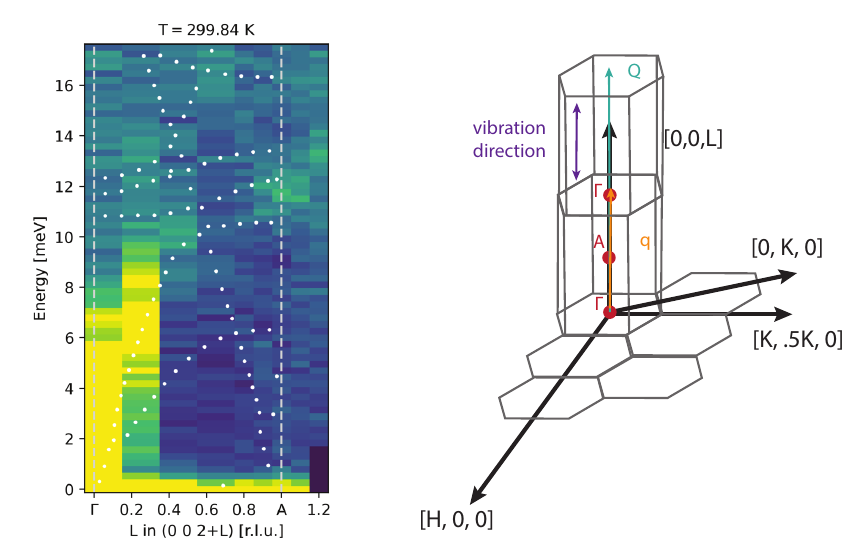}
 \caption{\textbf{Inelastic neutron scattering from transverse phonons.} Left, phonon bandstructure along the $\Gamma-A-\Gamma$ direction measured with triple axis spectroscopy along the 00L direction. White dotted lines are overlayed from the AIMD calculations of the phonon bandstructure at this temperature.  Right, schematic of scattering geometry with neutron momentum Q, phonon wavevector q, and the resulting vibration direction indicated with teal, orange, and purple arrows respectively In contrast to the INS measurements shown in main text, the vibration direction of these phonons is transverse relative to the kagomé plane. }
 \label{supfig:TAX phonon}
\end{figure}

\begin{figure}[ht!]
 \includegraphics[width=1.0\columnwidth]{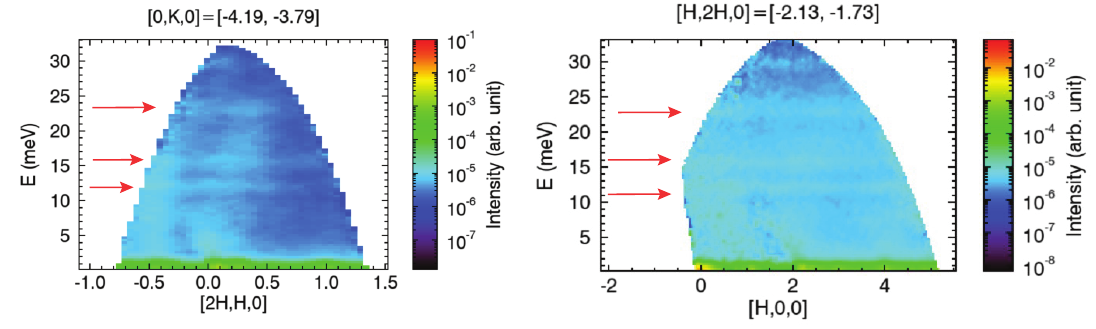}
 \caption{\textbf{Phonon bandstructure from time-of-flight measurements.} Inelastic neutron scattering measurements of the phonon band structure using the time-of-of flight method at the Spallation Neutron Source reveal flat phonon bands, indicated by the orange arrows. Left and right panels depict two different directions in reciprocal space. The phonons are flat across multiple Brillouin zones.}
 \label{supfig:ARCS phonon}
\end{figure}

\section*{Supplementary Note 8: Magnetostriction}\label{sec:magnetostriction}

Magnetostiction experiments were performed with a commercially available mini-dilatometer \cite{Kuchler2023} which is compatible with the sample environment provided by the Quantum Design PPMS Dynacool. This device can be rotated relative to the applied magnetic field to find the magnetostriction for field applied along specific crystallographic axes. The sample was mounted such that the change in length was measured along the $[210]$ direction for fields applied along the $[110]$ and $[010]$ directions. The resulting dilation at measured at different temperatures can be fit based on the phenomenological model derived above such that 

$$\Delta L /L \propto a + b H^2 + c H^4$$

where a, b, and c are constants and H is the value of applied magnetic field. Fits to this model at higher magnetic field values are shown in red in SI Figure 11. For each direction, the extracted coefficient to the $H^2$ term are plotted as a function of temperature in SI Figure 11. At T$=235$ K there is a kink in the coefficient's behavior, and below this temperature b is negative for the field along the [010] direction but positive for field along the [110] direction. We also construct a proxy value $\eta$ for the nematicity based on these coefficients

$$\eta = \frac{b_{[010]} - b_{[110]}}{b_{[010]} + b_{[110]}}$$

Which is plotted in red. Note that near $T^* = 225K$, this value approaches diverges but approaches zero, at temperatures away from this value. There is another divergence near T = 240 K that arises when the denominator of $\eta$ diverges.

\begin{figure}[ht!]
 \includegraphics[width=1.0\columnwidth]{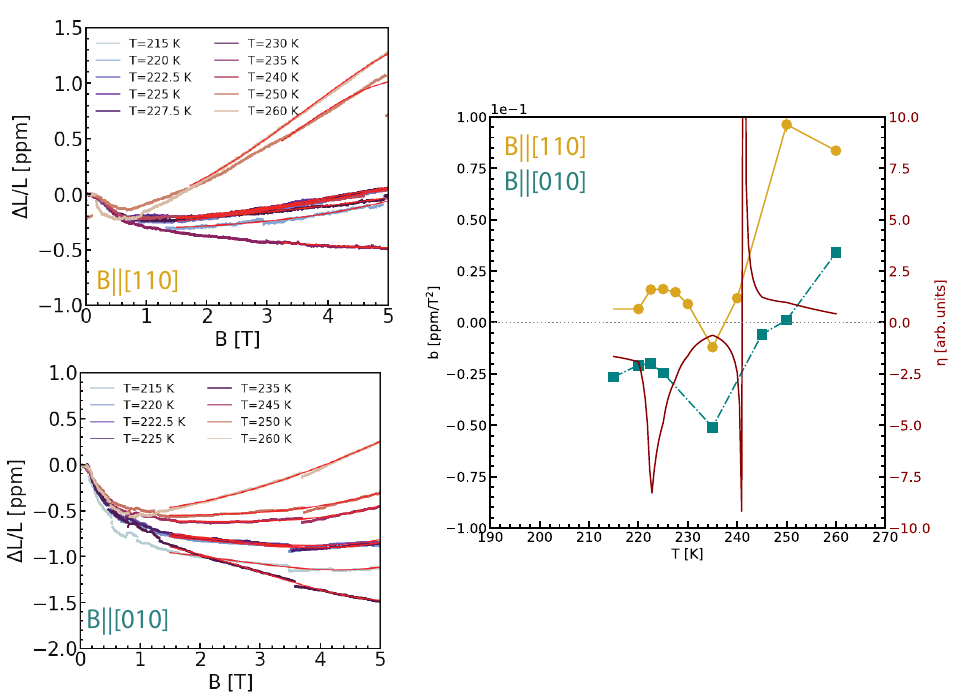}
 \caption{\textbf{Magnetostriction scaling.} Right, magnetostriction along the [210] direction for fields applied along the [110] and [010] directions top and bottom, respectively for different temperatures through the anomaly. Solid red lines indicate a fit to $\Delta L/L = a+b*B^2+c*B^4$ as predicted from first principles. Right, coefficient of the square term $b$ for each field direction in yellow and teal respectively along with a nematicity parameter $\eta = (b_{[110]}-b_{[010]})/(b_{[110]}+b_{[010]})$ in dark red. }
 \label{supfig:MS}
\end{figure}

\section*{Supplementary Note 9: Neutron Larmor Diffraction}\label{sec:larmor-diffraction}

High resolution neutron Larmor diffraction was used to measure the lattice spacings along different crystallographic directions\cite{Rekveldt_2001, Wang2018} at the HB-1 beamline at the High Flux Isotope Reactor (HFIR) at Oak Ridge National Laboratory \cite{Li2017}. This technique depends on the change in Larmor phase $\phi$ of polarized neutrons subjected to a magnetic field $B$ before and after scattering from a single-crystal sample. The measured Larmor phase is 

$$\phi = \frac{\gamma_N m d BL}{2\pi h}$$

Where $\gamma_N$ is the gyromagnetic ratio of the neutron, $m$ is the neutron mass, $d$ is the lattice spacing, $B$ is the field strength and $L$ is path length of the field. In this way, small changes in the lattice spacing $d$ are detectable by a change in Larmor phase $\phi$. 

Coaligned samples were oriented in different scattering geometries in order to measure along the [201], [021], and [002] directions. 

One advantage of this technique is that a change in the total magnitude of $\phi$ can detect structural phase transitions \cite{Rekveldt_2001, Li2017} up to a resolution of $\Delta d/d = 1e-5$. In our measurements, there was not a significant change in the overall magnitude of $\phi$ at temperatures near T$^*= 225 $K as shown in SI figure 13. 

\begin{figure}[ht!]
 \includegraphics[width=1.0\columnwidth]{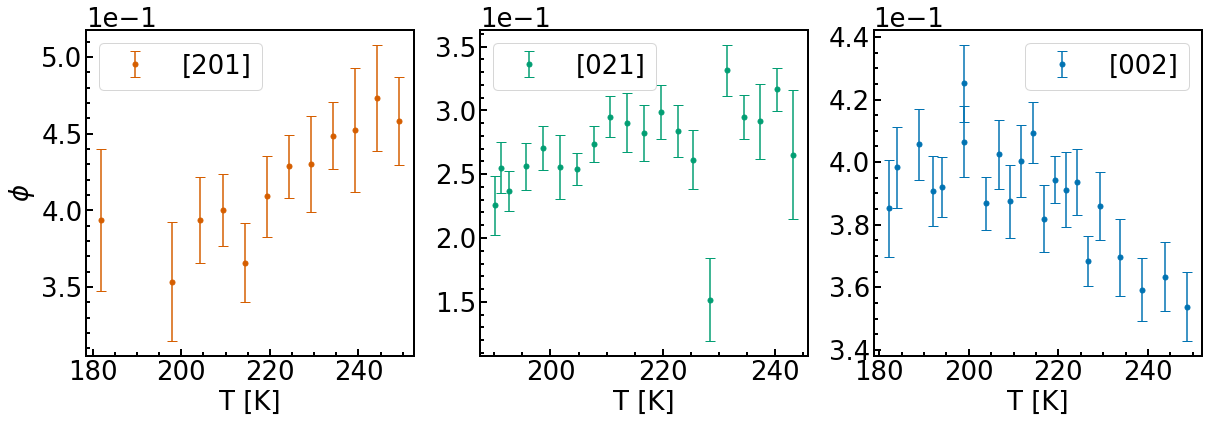}
 \caption{\textbf{Larmor diffraction amplitude.} Total neutron Larmor phase amplitude at different temperatures through the nematic region for different crystallographic orientations. The lack of sharp change in the Larmor phase amplitude suggests that a structural phase transition is not detected despite the change in lattice distortion.}
 \label{supfig:sampleimage-12}
\end{figure}

\section*{Supplementary Note 10: Thermodynamics and susceptibility of a kagomé lattice model}\label{sec:theory_micro}

Here we show that susceptibility of flat-band electrons in a kagomé lattice shows a pronounced peak at an intermediate temperature.
We follow the notation of \cite{liu2020_th}:
\begin{equation}
H({\bf k}) 
=
E_0+
    \begin{pmatrix}
    0&t_1+t_1 e^{-2 i k_x a}&t_2+t_2 e^{-i (kx + \sqrt{3} ky)a}\\
    t_1+t_1 e^{-2 i k_x a}&0&t_3+t_3 e^{-i (-kx + \sqrt{3} ky)a}\\
    t_2+t_2 e^{i (kx + \sqrt{3} ky)a}&t_3+t_3 e^{i (-kx + \sqrt{3} ky)a}&0
    \end{pmatrix},
    \label{eqsup:h0}
\end{equation}
where in the absence of distortions $t_1=t_2=t_3=t$.

Strain deforms the kagomé lattice, such that $t_{1,2,3}$ become non-equivalent \cite{liu2020_th}. For example, $\varepsilon_{xx}$ strain results in 
\begin{equation}
t_1= t, \; t_2= t + \alpha,\; t_3= t + \alpha,
\label{eqsup:hstr}
\end{equation}
where $\alpha\propto \varepsilon_{xx}$.

The action of $B_{1u}$ deformation is to bring three atoms closer to the center of hexagons, ultimately forming trimers (see Fig. 5 (c) of main text). This can be modeled by an increase/decrease in next-nearest-neighbor tunneling strengths as follows:
\begin{equation}
\delta H_{B_{1u}}({\bf k}) 
=
\delta
    \begin{pmatrix}
    0&e^{i (-k_x + \sqrt{3} k_y)}-e^{i (-k_x - \sqrt{3} k_y)}&e^{-2 i k_x}-e^{i (k_x - \sqrt{3} k_y)}\\
    e^{-i (-k_x + \sqrt{3} k_y)}-e^{-i (-k_x - \sqrt{3} k_y)}&0&e^{i (-k_x - \sqrt{3} k_y)}-e^{2 i k_x} \\
   e^{2 i k_x}-e^{-i (k_x - \sqrt{3} k_y)}&e^{-i (-k_x - \sqrt{3} k_y)}-e^{-2 i k_x} &0
    \end{pmatrix},
    \label{eqsup:hb1u}
\end{equation}

In Fig. 5 of the main text we show the band structure cut and density of states for (a) $t_1=t_2=t_3=-1$, (b) $t=-1, \alpha = 0.1$  (c) $t=-1, \delta=-0.2$. The results were calculated numerically, using 512x512 k-points in the Brillouin zone and Lorentzian energy smearing of each eigenstate with width $0.01 t$. For both distortions, the most dramatic effect is the smearing of the density of states peak due to the flat band.

\begin{figure}[ht!]
 \centering
 \includegraphics[width=\columnwidth]{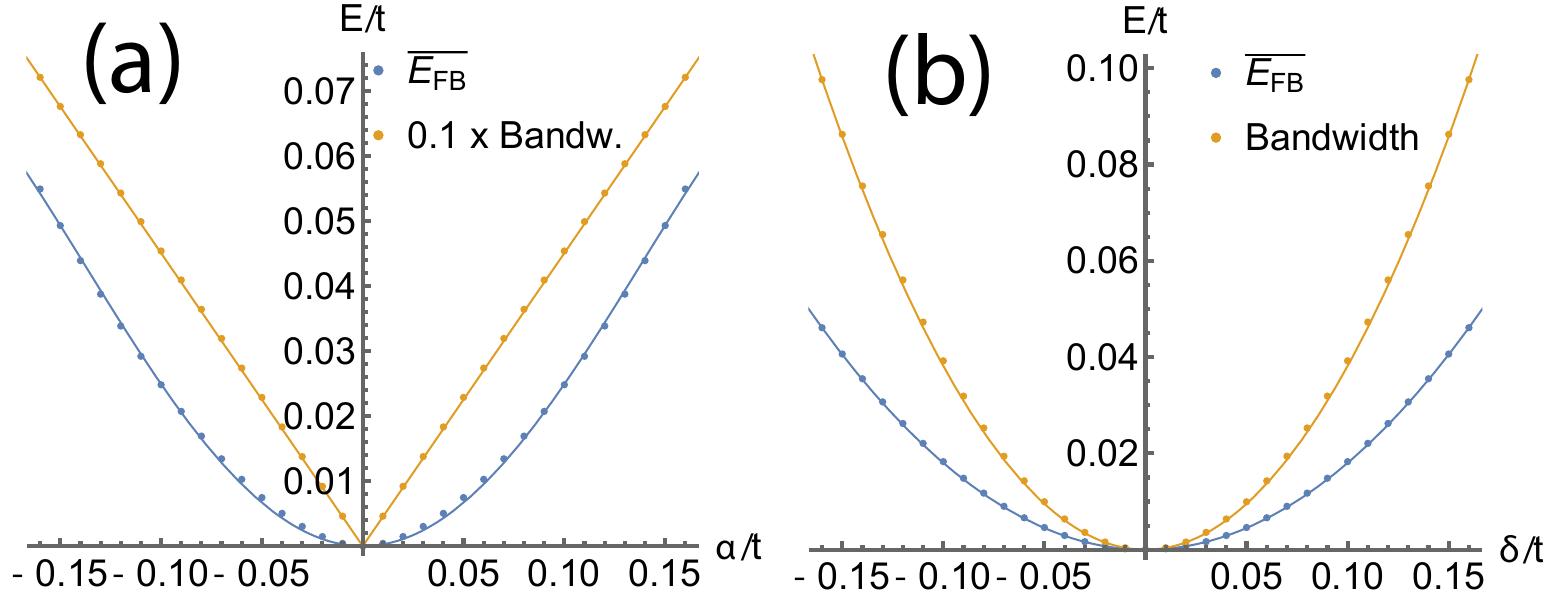}
 \caption{\textbf{Change of the flat band parameters for distorted kagomé model} (a) Effect of strain (b) $B_{1u}$ distortion. Lines are polynomial fits to the numerical points.}
 \label{supfig:theor1}
\end{figure}

In Fig. \ref{supfig:theor1} we show the dependence of the top band's width and its averaged energy as a function of $\alpha$ or $\delta$. For strain ($\alpha$), there is a dramatic increase of the flat band width, that is linear in $\alpha$, while $B_{1u}$ distortion leads to a quadratic increase. In both cases the average energy of the flat band shifts, but for strain the dependence is less well described by quadratic; the fit shown in Fig. \ref{supfig:theor1} (a) is of the form $a x^2+b x^4$, while it is purely quadratic in (b).

Let us now consider the temperature dependence of the system's susceptibility to such distortions. We compute numerically the free energy with and without the distortion including full band structure of the model, Eq. \eqref{eqsup:h0},\eqref{eqsup:hb1u}. 

\begin{equation}
\begin{gathered}
\chi_{\alpha} 
=
-\frac{\partial ^2 F}{\partial \alpha^2}
;
\;\;\;
\chi_{\delta} 
=
-\frac{\partial ^2 F}{\partial \delta^2 }
\\
F (T,\alpha,\delta)= - 2 T \sum_{\bf k}\log \left[
\prod_{i=1}^3 (1+e^{-E_i({\bf k},\alpha,\delta)/T})
\right],
\end{gathered}
\label{eqsup:susc}
\end{equation}
where $E_i({\bf k})$ are the eigenvalues for the distorted kagomé models described above.

\begin{figure}[ht!]
 \centering
 \includegraphics[width=\columnwidth]{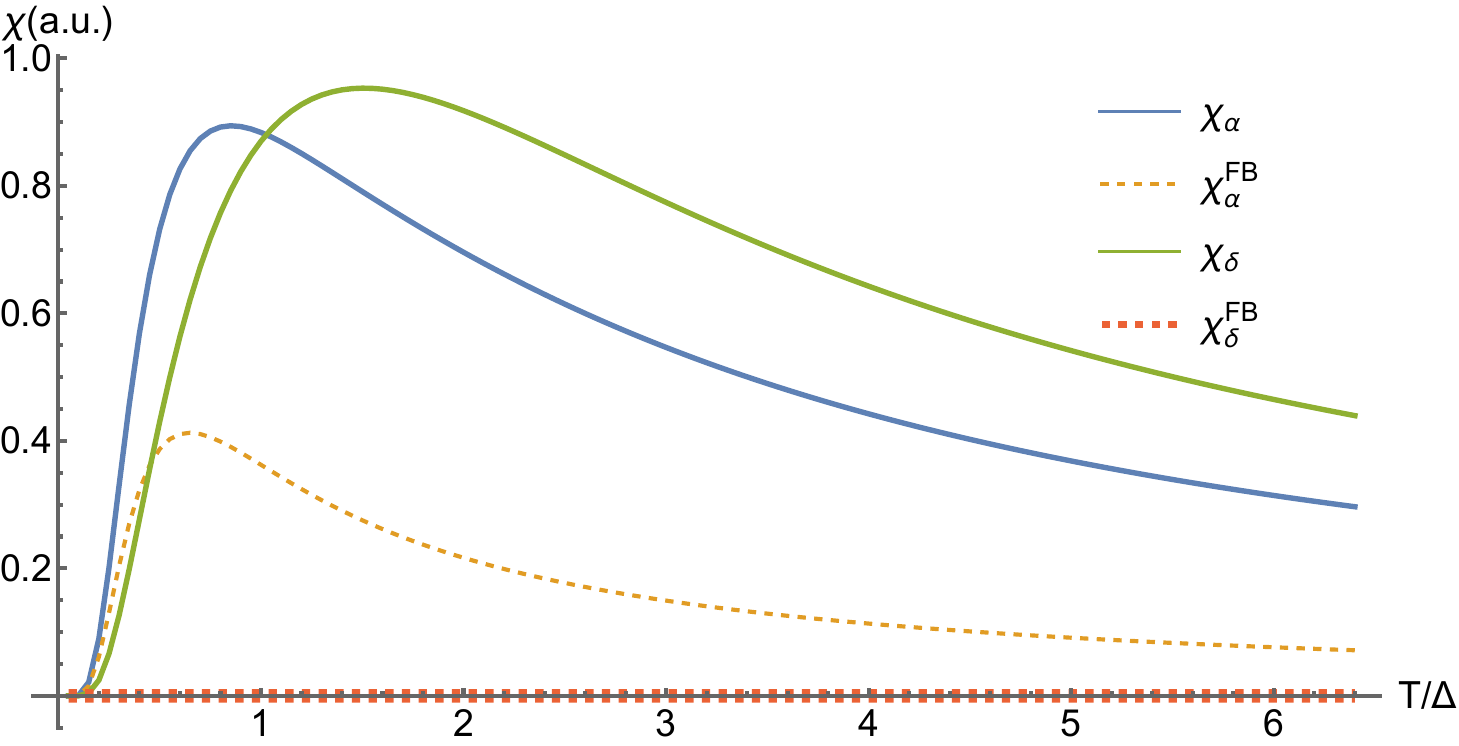}
 \caption{\textbf{Kagomé model susceptibilities.} Solid lines are the susceptibilities, Eq. \eqref{eqsup:susc} with respect to strain (blue) and $B_{1u}$ distortion (green) for $\Delta=-t=1$ corresponding to $E_0=3$. Dashed lines represent the contribution of the flat band broadening to susceptibility, Eq. \eqref{eqsup:chifb}.}
 \label{supfig:theor2}
\end{figure}

Fig. \ref{supfig:theor2} shows the susceptibility calculated numerically on a 128x128 momentum grid and with the derivative approximated by $\frac{\partial ^2 F}{\partial x^2} \approx [F(x)+F(-x)-2F(0)]/(2x^2)$ taken for $x \equiv \delta =0.01$ and $x \equiv \alpha =0.01$. To analyze the results we further separate the contribution arising from the flat band acquiring a finite bandwidth due to distortion, see Fig. \ref{supfig:theor1}. It is equal to:
\begin{equation}
\begin{gathered}
\chi_{\alpha}^{FB} 
=
-\frac{\partial ^2 F^{FB}}{\partial \alpha^2}
;
\;\;\;
\chi_{\delta}^{FB}  
=
-\frac{\partial ^2 F^{FB}}{\partial \delta^2 },
\\
F (T,\alpha,\delta)= - 2 T \sum_{\bf k}\log \left[
 (1+e^{-\frac{E_{FB}({\bf k},\alpha,\delta) - \overline{E}_{FB}(\alpha,\delta)}{T}})
\right].
\end{gathered}
\label{eqsup:chifb}
\end{equation}

One observes that both susceptibilities have a peak at a finite $T$ smaller then the gap value. For strain, a significant part of susceptibility can be attributed to the increasing dispersion of the flat band. The rest is due to shifts in average position of the bands and in principle should be less universal. Note that the flat band broadening does not contribute to the $B_{1u}$ susceptibility.

To capture the response due to broadening of the flat band that is expected to be more universal one can consider the 
following simplified model. For non-zero $\alpha$, we approximate the density of states as a "box" with a finite width equal to $c \alpha$. Then, the free energy per unit cell takes then the form:

\begin{equation}
\begin{gathered}
    F_{mod} = -T \int_{-\Delta - c\alpha/2}^{-\Delta + c\alpha/2}\frac{d\varepsilon}{c \alpha}  \log[1+e^{-\frac{\varepsilon}{T}}]
    \approx
    -T \log[1+e^{\frac{\Delta}{T}}]
    \\
    -\frac{e^{\frac{\Delta}{T}} }{24 T (1 + e^{\frac{\Delta}{T}})^2}c^2\alpha^2
+O(\alpha^4)
\end{gathered}
\end{equation}
From the above one extracts the expression for susceptibility given in the main text.

\section*{Supplementary Note 11: Specific heat for phenomenological model}\label{sec:theory_pheno}

The specific heat obtained from the free energy given in the main text is:

\begin{equation}
\delta C = -T \frac{\partial^2 F}{\partial T^2}
\approx
\frac{2T^*(G_\Phi(T^*) \vec{\sigma}^2-2G\vec{\kappa}\cdot \vec{\sigma})}{|G_\Phi''(T^*)| G_\varepsilon} 
\frac{3(T-T^*)^2 - T'^2 }{(T-T^*)^2 + T'^2},
\end{equation}
where we denote $T'=\frac{2} {|G_\Phi''(T^*)|}[G^2/G_\varepsilon-G_\Phi(T^*)]$.

We can parameterize the result as $\delta C = C_0 (1+a \kappa) \frac{3(T-T^*)^2 - T'^2 }{(T-T^*)^2 + T'^2}$

The function plotted in Fig. 5 (e) of the main text is $x+10^{-8}A\frac{3 (-1 + x)^2 - a^2}{a^2 + (-1 + x)^2}^3$ for $a=0.045$ and $A=1,9,25,64,100$, roughly mimicking a 100-fold increase of $\kappa$ (10-fold increase in magnetic field).

\section*{Supplementary Note 12: Effect on $B_{1u}$ phonons for phenomenological model}\label{sec:theory_pheno_2}

The dynamics of coupled particle-hole excitations in a non-symmetric channel and phonons can be described using the following equations \cite{volkov_excitonic}:

\begin{equation}
\begin{gathered}
\{\partial_t+\Omega_e(T) \} \varphi +V \eta =0,
\\
\{\partial^2_t+2\gamma(T) \partial_t+ \omega_{p}^2(T) \}\eta + V \varphi =0,
\end{gathered}
\label{eq:dyn}
\end{equation}
where $\varphi$ - is the collective coordinate for particle-hole excitations in $B_{1u}$ channel, $\Omega_e(T) \propto \chi^{-1}_{B_{1u}}(T) =  G_0^{B1u} - \chi_{\delta}(T)$ which has a minimum at a finite $T$ and $\eta$ - is the phonon coordinate, $\gamma(T)$ and $\omega_{p}(T)$ being phonon damping and frequency, respectively.

For $V\ll\omega_p\ll\Omega_e$, solving the first equation and inserting the result in the second one, one obtains a shift in phonon energy equal to $-\frac{V^2}{\Omega_e(T)} \propto -\frac{V^2}{G_0^{B1u}}-\frac{V^2}{(G_0^{B1u})^2}\chi_{\delta}(T)$ for  $\chi_{\delta}(T)\ll G_0^{B1u}$. The second term will have a strong peak at an intermediate temperature.
